\documentclass[acmlarge,nonacm]{acmart}
\def\BibTeX{{\rm B\kern-.05em{\sc i\kern-.025em b}\kern-.08emT\kern-.1667em\lower.7ex\hbox{E}\kern-.125emX}}
\usepackage{booktabs}
\usepackage[shortend,ruled,linesnumbered]{algorithm2e}
\usepackage[titletoc]{appendix}

\setcopyright{acmlicensed}
\acmJournal{IMWUT}
\acmYear{2021} \acmVolume{5} \acmNumber{4} \acmArticle{184}
\acmMonth{12} \acmPrice{15.00}\acmDOI{10.1145/3494993}

\usepackage{amsthm}
\usepackage{subfigure}
\usepackage{multirow}
\usepackage{diagbox}
\usepackage{color}
\usepackage{colortbl}
\usepackage[normalem]{ulem}

\newcommand{\ign}[1]{}
\usepackage{tabu}
\newcommand{\para}[1]{{\vspace{4pt} \bf \noindent #1 \hspace{10pt}}}
\newenvironment{packed_itemize}{
\begin{list}{\labelitemi}{\leftmargin=1.5em}
  \setlength{\itemsep}{1pt}
  \setlength{\parskip}{0pt}
  \setlength{\parsep}{0pt}
  \setlength{\headsep}{0pt}
  \setlength{\topskip}{0pt}
  \setlength{\topmargin}{0pt}
  \setlength{\topsep}{0pt}
  \setlength{\partopsep}{0pt}
}{\end{list}}
\usepackage{epstopdf}
\newtheorem{mdefinition}{Definition}

\begin{document}
\title{Spatio-Temporal Urban Knowledge Graph Enabled Mobility Prediction}

\author{Huandong Wang}
\email{wanghuandong@tsinghua.edu.cn}
\authornote{The first two authors have equal contribution.}
\author{Qiaohong Yu}
\email{yuqiaohong20@mails.tsinghua.edu.cn}
\authornotemark[1]
\author{Yu~Liu}
\email{liuyu18@mails.tsinghua.edu.cn}
\author{Depeng~Jin}
\email{jindp@tsinghua.edu.cn}
\author{Yong~Li}
\email{liyong07@tsinghua.edu.cn}
\affiliation{%
  \institution{Beijing National Research Center for Information Science and Technology (BNRist), Department of Electronic Engineering, Tsinghua University, China}
  \country{China}
  }

\begin{abstract}
With the rapid development of the mobile communication technology, mobile trajectories of humans are massively collected by Internet service providers (ISPs) and application service providers (ASPs). On the other hand, the rising paradigm of knowledge graph (KG) provides us a promising solution to extract structured ``knowledge'' from massive trajectory data. In this paper, we focus on modeling users' spatio-temporal mobility patterns based on knowledge graph techniques, and predicting users' future movement based on the ``knowledge'' extracted from multiple sources in a cohesive manner. Specifically, we propose a new type of knowledge graph, i.e., spatio-temporal urban knowledge graph ({\em STKG}), where mobility trajectories, category information of venues, and temporal information are jointly modeled by the facts with different relation types in {\em STKG}. The mobility prediction problem is converted to the knowledge graph completion problem in {\em STKG}. Further, a complex embedding model with elaborately designed scoring functions is proposed to measure the plausibility of facts in {\em STKG} to solve the knowledge graph completion problem, which considers temporal dynamics of the mobility patterns and utilizes PoI categories as the auxiliary information and background knowledge. Extensive evaluations confirm the high accuracy of our model in predicting users' mobility, i.e., improving the accuracy by 5.04\% compared with the state-of-the-art algorithms. In addition, PoI categories as the background knowledge and auxiliary information are confirmed to be helpful by improving the performance by 3.85\% in terms of accuracy. Additionally, experiments show that our proposed method is time-efficient by reducing the computational time by over 43.12\% compared with existing methods.
\end{abstract}

\keywords{Knowledge Graph, Mobility Prediction, Spatio-Temporal Pattern}

\maketitle

\section{Introduction}\label{sec:intro}

With the rapid development of the mobile communication technology and the popularization of mobile devices, an explosive amount of mobile trajectories are collected by Internet service providers (ISPs) and application service providers (ASPs). Both of them have strong motivations to develop a high-performance mobility prediction model. For example, in mobile networks, by predicting the future movements of users, network operators can reserve or allocate resources in advance to optimize the quality of experience of users~\cite{nadembega2014destination,ni2011mpbc,soh2003qos}. For ASPs, accurately predicting user movement is also instrumental for providing intelligent services, especially for location-based applications~\cite{ying2012urban, zheng2010collaborative}. However, predicting users' mobility is also challenging due to the large number of spatial venues with different urban functions, diverse spatio-temporal mobility patterns, and numerous influencing factors of the users' future mobility.

At the same time, the rising paradigm of knowledge graph (KG) provides us a promising solution to extract structured ``knowledge'' from massive data. A number of applications including language understanding and recommendation systems based on knowledge graph techniques have shown great success~\cite{Sun2018Recurrent,liu2020kEnabling}. Thus, it shows a great potential to utilize knowledge graph techniques to model the complex relationship between spatial venues with different urban functions and users with diverse spatio-temporal mobility patterns. Then, an interesting research question is whether we can extract structured ``knowledge'' from the massive spatio-temporal trajectory data based on knowledge graph techniques. What's more,  can we incorporate ``knowledge'' from multiple sources, e.g., category information of venues obtained from the map services, to predict users' future movement with better performance?

A number of approaches have been proposed to utilize knowledge graphs to better model user behaviors~\cite{Wang2020Incremental,Wang2018DKN}. However, most of them directly enhance existing user models by using the embeddings of entities learned from KG as the initial features, which fails to incorporate ``knowledge'' from multiple sources in a cohesive manner. What's more, the ``knowledge'' involved in the user traces is ignored in these methods. For example, based on the trajectory data, we can extract the spatio-temporal mobility patterns of users, which are definitely important knowledge and useful for a number of applications including user profiling~\cite{xu2018detecting}, spatio-temporal aware recommendation~\cite{yuan2017pred}, etc.

In this paper, we focus on modeling users' spatio-temporal mobility patterns based on knowledge graph techniques, and predicting users' future movement based on the ``knowledge'' extracted from multiple sources in a cohesive manner. By elaborately defining a data schema, we construct our proposed spatio-temporal urban knowledge graph ({\em STKG}), where users, spatial venues, their categories, and time are regarded as entities in the {\em STKG} and ``knowledge'' from different sources are characterized by the facts of different types of relations between them. Further, the mobility prediction problem is converted to the knowledge graph completion problem in the {\em STKG}. We further propose a multi-ary embedding model in the complex vector space to solve the knowledge graph completion problem, where entities and relations are jointly mapped into a low-dimensional complex space with the plausibility of the facts between them preserved. In this process, multi-level categories of spatial venues are jointly considered in a uniform manner, which characterizes their urban functions. At the same time, user features are learned correlated with their diverse spatio-temporal mobility patterns. What's more, we define different types of relations to characterize the spatio-temproal patterns correlated with different combinations of the influencing factors, and implement extensive experiments to evaluate their influence on the performance of the mobility prediction. In summary, our paper makes the following contributions:
\begin{packed_itemize}
\item This is the first work to solve the mobility prediction problem by converting it to a knowledge graph completion problem in the new proposed spatio-temporal urban knowledge graph ({\em STKG}), where massive mobility trajectories, category information of venues, and temporal information are jointly modeled in a cohesive manner.
\item We propose a complex embedding model for entities and relations in the {\em STKG} with elaborately designed scoring functions to model the facts between them, which considers temporal dynamics of the mobility patterns and auxiliary information with different forms describing the historically visited venues and their categories.
\item Extensive evaluations confirm the high accuracy of our model in predicting users' mobility. Specifically, our proposed model improves the accuracy of user mobility prediction by 5.04\% compared with the state-of-the-art algorithms. What's more, background knowledge is confirmed to be helpful, i.e., by combining the auxiliary information and the affiliation relations based on the PoI category information, the prediction performance of our proposed model is improved by 3.85\% in terms of accuracy. 
In addition, experiments show that our proposed method is time-efficient by reducing the computational time by over 43.12\% compared with existing methods.
\end{packed_itemize}

The rest of the paper is structured as follows. In Section~\ref{sec:model}, we present the mathematical model and formulate the investigated problem. In Section~\ref{sec:method}, we introduce the methodology to solve the mobility prediction problem based on our proposed spatio-temporal urban knowledge graph. In Section~\ref{sec:evaluation}, we evaluate the performance of our proposed algorithms compared with existing algorithms. Finally, after discussing related work in Section~\ref{sec:relatedwork}, we summarize our main findings in Section~\ref{sec:conclusion}.

\begin{table}[t]
\begin{center}
\caption{Notations and descriptions.}\label{tab:notation}
\begin{tabu}{|p{2.1cm}<{\centering}|[1.02pt]p{13cm}|}
\tabucline[1.02pt]{-}
\textbf{Notation} & \textbf{Description} \\ \hline
 $\mathcal{G}$ & The spatio-temporal urban knowledge graph.\\
 $\mathcal{G}^V$ & The spatio-temporal urban knowledge graph with only spatio-temporal mobility pattern relations.\\
 $\mathcal{G}^{Ci}$ & The spatio-temporal urban knowledge graph with only $i$-th level PoI categories affiliation relations.\\
 $\mathcal{G}^{C}$ & The spatio-temporal urban knowledge graph with all PoI categories affiliation relations, i.e., $\cup^3_{i=1}\mathcal{G}^{Ci}$.\\
 $\mathcal{U}$ & The set of users.\\
 $\mathcal{T}$ & The set of time bins.\\
 $\mathcal{P}$ & The set of PoIs.\\
 $\mathcal{R}$ & The set of relation types.\\
 $\mathcal{C}_1,\mathcal{C}_2,\mathcal{C}_3$ & The set of three level categories of PoIs.\\
 $\mathcal{C}$ & The set of all categories, i.e., $\mathcal{C}=\{\mathcal{C}_1,\mathcal{C}_2,\mathcal{C}_3\}.$\\
 $x=(u,l,t)$ & The mobility record indicating that user $u$ visits location $l$ at time $t$.\\
 $tr^u$ & The mobility trajectory of user $u\in\mathcal{U}$.\\
 $r_V$ & Spatio-temporal mobility pattern relation.\\
 $r_{C1},r_{C2},r_{C3}$ & Affiliation relation of three level categories, respectively.\\
 $\alpha$ & The fraction of time-modulated dimensions in the embedding vector $r_V^E(t,A)$.\\
 $\beta$ & The parameter to balance the influence of the loss of different types of relations.\\
\tabucline[1.02pt]{-}
\end{tabu}
\label{tab:symbal}
\end{center}
\end{table}

\section{Preliminaries and System Overview}\label{sec:model}

In this section, we provide preliminaries of our paper by introducing fundamental concepts and formally defining the mobility prediction problem that we investigate. In addition, we also provide an overview of our proposed knowledge graph based mobility prediction system. Table~\ref{tab:notation} shows the meanings of key notations used in this paper.

\subsection{Definitions}

\begin{mdefinition}[Point of Interest]
A point of interest (PoI) is a spatial venue with a specific urban function, e.g., a restaurant or an office building. In addition, each PoI is associated with several PoI categories which contain PoIs with the same urban function. 
\end{mdefinition}

We denote the set of PoIs as $\mathcal{P}$. In addition, we assume there exist three levels of PoI categories, i.e., fine-level categories, mid-level categories, and coarse-level categories. We denote the set of categories with these three different levels as $\mathcal{C}_1$, $\mathcal{C}_2$, and $\mathcal{C}_3$, respectively.

\begin{mdefinition}[Mobility Record]
Each mobility record is defined as 3-tuple $x=(u,p,t)$, where $u$ is the identifier of the user or the device, $t$ is the timestamp, and $p$ is the PoI visited by user $u$ at time $t$.
\end{mdefinition}

In order to process timestamps with different resolutions in different datasets in a unified manner, we discretize timestamps into fixed-length time bins. Following the widely used preprocessing strategy of timestamps in existing approaches~\cite{zhuang2017understanding,xu2018detecting}, we divide one day into 48 half-hour time bins, and we further distinguish time bins belonging to the working day and non-working day. Thus, there are totally 96 time bins. We further denote the set of time bins as $\mathcal{T}$.

\begin{figure*}[t]
\centering
\subfigure{\includegraphics[width=.75\textwidth]{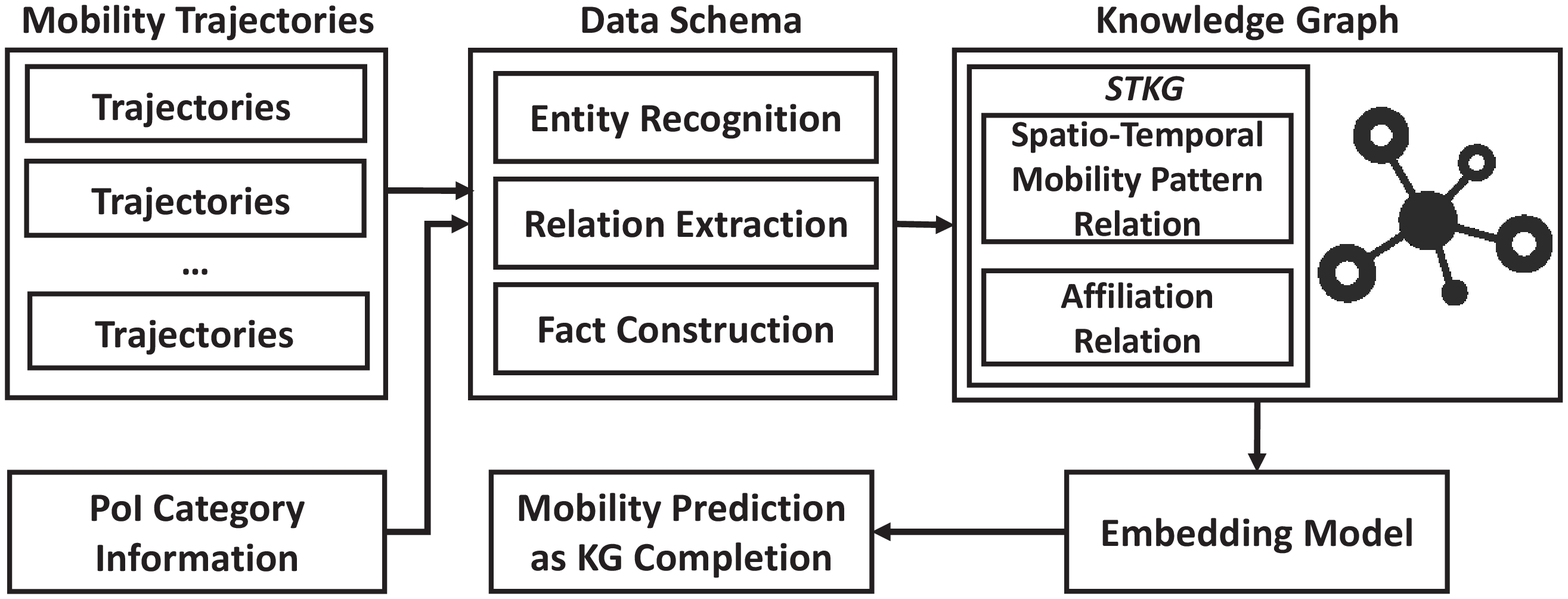}}
\caption{The workflow of our proposed system.} \label{fig:system}
\end{figure*}

\begin{mdefinition}[Mobility Trajectory]
For an arbitrary user or device $u$, its mobility trajectory is defined by a sequence of ordered spatio-temporal mobility records as $tr^u=\{x^u_1,x^u_2,...,x^u_n\}$, where $x^u_i=(u,p^u_i,t^u_i)$ is a mobility record. In addition, for arbitrary $1\leq i\leq j\leq n$, we have $t^u_i\leq t^u_j$.
\end{mdefinition}

\subsection{Problem Formulation}

Based on the above-defined notations and concepts, the mobility prediction problem that we investigate can be expressed as follows:
\begin{mdefinition}[Mobility Prediction Problem]\label{def:POI_distribution}
Given the historical mobility trajectories $\{tr^u\}_{u\in \mathcal{U}}$ and the target time $\{t^{u}_\star\}_{u\in \mathcal{U}}$, estimate the PoI $p^{u}_\star$ of each user $u\in\mathcal{U}$ at time $t^{u}_\star$. In addition, auxiliary information can be utilized in this process, e.g., category information of PoIs.
\end{mdefinition}

\subsection{System Overview}

We show the workflow of our proposed system in Figure~\ref{fig:system}. As we can observe, based on the collected mobility trajectory data and PoI category information, we first utilize a pre-defined data schema to extract entities, relations, and facts to construct the spatio-temporal urban knowledge graph ({\em STKG}). Then, we propose an embedding model to learn the feature vectors of all entities in our {\em STKG}. Finally, based on the learned embedding vectors, we implement mobility prediction, which is converted to the knowledge graph completion problem in the constructed {\em STKG}.
\section{Methodology}\label{sec:method}

In this section, we present our mobility prediction method based on knowledge graphs. We first define the concept of the spatio-temporal urban knowledge graph ({\em STKG}), and introduce how to construct it based on mobility trajectories and PoI category information and how to convert the mobility prediction problem to a knowledge graph completion problem in the {\em STKG}. Then, we propose a representation learning model to extract the features of entities and relations from the facts in the observed {\em STKG}. Following the description about how to optimize the model parameters, we present the algorithm details about how to implement mobility prediction based on the {\em STKG}, and analyze the algorithms in terms of computational complexity.

\subsection{Knowledge Graph Construction}

We first introduce how we define the {\em STKG}, and extract relations from trajectories and PoI category information. Specifically, the definition of {\em STKG} is given as follows:

\begin{mdefinition}[Spatio-Temporal Urban Knowledge Graph ({\em STKG})]\label{def:stkg}
The spatio-temporal urban knowledge graph is defined as a graph $\mathcal{G}=(\mathcal{E},\mathcal{R},\mathcal{F})$, where $\mathcal{E}$ is the set of spatio-temporal entities including users, time units, PoIs, and their categories.
$\mathcal{R}$ represents the set of relation types between entities,
and $\mathcal{F}$ represents the set of facts corresponding to different relation types. $\mathcal{R}$ and $\mathcal{F}$ jointly describe the interrelation between the spatio-temporal entities in $\mathcal{E}$.
More specifically, there are two types of relations in $\mathcal{G}$. The first type is the spatio-temporal mobility pattern relation, which describes the fact that a user visits some PoI at a certain time. The second type is the affiliation relation between PoI and PoI categories, which describes the fact that a PoI belongs to a PoI category.
\end{mdefinition}

As we can observe from Definition~\ref{def:stkg}, our defined {\em STKG} contains two types of relations, where the spatio-temporal mobility pattern relation is constructed based on the historical mobility trajectories of users and the affiliation relation is constructed based on the PoI category information. We further formally define these two relations as follows:

\begin{mdefinition}[Spatio-Temporal Mobility Pattern Relation ({\em STMPR})]
If the fact that the user $u\in\mathcal{U}$ visited the PoI $p\in\mathcal{P}$ at time $t\in\mathcal{T}$ is true, the spatio-temporal mobility pattern relation $r_V$ is defined to exist between $u$ and $p$.
Further, the fact is denoted by a tuple $(u,r_V,p,t,A)$, where $A$ is the auxiliary information describing the spatio-temporal context of this fact.
\end{mdefinition}

In fact, the time $t$ is regarded as auxiliary information of the relation in a number of existing studies~\cite{galkin2020message,jiang2016towards}. However, the time information plays a critical role in the spatio-temporal mobility pattern relation, i.e., the mobility pattern is valid only at certain points in time. Thus, we regard the time as an additional entity in the relation, while $A$ incorporates other mobility-related entities, including the most recently visited PoI, the corresponding category, etc. Specifically, we define five types of spatio-temporal mobility pattern relations with different auxiliary information in Table~\ref{tab:stmpr}.

\begin{table}[t]
\vspace{-0.3cm}
\begin{small}
\begin{center}
\caption{Spatio-temporal mobility pattern relations (STMPR) with different auxiliary information.}\label{tab:patterns}
\vspace{-0.4cm}
\begin{tabu}{|p{2.1cm}<{\centering}|[1.02pt]p{13cm}|}
\tabucline[1.02pt]{-}
\textbf{Relation} & \textbf{Corresponding Auxiliary Information} \\ \hline
 $r_{V0}$ & $A=\emptyset$.\\
 $r_{V1}$ & $A=\{p_s\}$, where $p_s\in\mathcal{P}$ is the most recently visited PoI.\\
 $r_{V2}$ & $A=\{c_{s1}\}$, where $c_{s1}\in\mathcal{C}_1$ is the fine-level category of the most recently visited PoI.\\
 $r_{V3}$ & $A=\{c_{s2}\}$, where $c_{s2}\in\mathcal{C}_2$ is the mid-level category of the most recently visited PoI.\\
 $r_{V4}$ & $A=\{c_{s3}\}$, where $c_{s3}\in\mathcal{C}_3$ is the coarse-level category of the most recently visited PoI.\\
\tabucline[1.02pt]{-}
\end{tabu}
\label{tab:stmpr}
\end{center}
\end{small}
\vspace{-0.3cm}
\end{table}

The facts of spatio-temporal mobility pattern relations can be constructed by attaching the corresponding auxiliary information $A$ to each mobility record $x=\{u,p,t\}$. Specifically, based on the considered auxiliary information summarized in Table~\ref{tab:patterns}, for each mobility record $x=\{u,p,t\}$, we only need to find the most recently visited PoI $p_s$ of the user from his/her trajectory $tr^u$. Then, use $p_s$ or the category $c\in\mathcal{C}_i$ which $p_s$ belongs to as the auxiliary information $A$. 

\begin{mdefinition}[Affiliation Relation]
For each $c_1\in\mathcal{C}_1$ and an arbitrary PoI $p$, if the fact that $c_1$ is the fine-level category of $p$ is true, the affiliation relation $r_{C1}$ is defined to exist between $c_1$ and $p$, and this fact is denoted by a 3-tuple $(p,r_{C1},c_1)$. Similarly, for $c_2\in\mathcal{C}_2$ and $c_3\in\mathcal{C}_3$, the facts of that $c_2$ and $c_3$ are the mid-level and coarse-level category of $p$ are defined as $(p,r_{C2},c_2)$ and $(p,r_{C3},c_3)$, respectively.
\end{mdefinition}

The facts of affiliation relations can be simply constructed by going through PoIs and finding their affiliated PoI categories with different levels. For the sake of simplicity, we denote the sub-{\em STKG} with only spatio-temporal mobility pattern relations and affiliation relations of $i$-th level categories as $\mathcal{G}^V$ and $\mathcal{G}^{Ci}$, respectively.

Actually, another type of affiliation relation exists between PoI categories of different levels, defined as the category affiliation relation. The category affiliation relations describe the fact that a fine-level PoI category belongs to a mid-level PoI category, or a mid-level PoI category belongs to a coarse-level PoI category. We also denote the sub-{\em STKG} with only the category affiliation relations as $\mathcal{G}^{C'}$.  
Note that the category affiliation relation does not have a direct impact on human mobility, and thus their influence on the prediction performance is less than the spatial-temporal mobility pattern relation and the affiliation relation, which will be verified in Section~\ref{sec:evaluation}.
Thus, we mainly consider the two key relations of
the spatio-temporal mobility pattern relation and the affiliation relation in our constructed {\em STKG}.

\para{Modeling the mobility prediction problem as a KG completion problem.} Given the definition of spatio-temporal urban knowledge graph above, the mobility prediction problem can be transformed into a knowledge graph completion problem, which is formally defined as follows:

\begin{mdefinition}[Spatio-Temporal Urban Knowledge Graph Completion Problem]
Given the observed spatio-temporal urban knowledge graph $G$,
for each $u\in\mathcal{U}$, the target time $t^{u}_\star$, the corresponding auxiliary information $A$, and a PoI $p\in\mathcal{P}$, infer whether the fact $(u,r_V,p,t^{u}_\star)$ is true.
\end{mdefinition}

Based on the definitions above,
by enumerating all possible candidate locations $p\in\mathcal{P}$ and ranking them based on the plausibility of their corresponding facts, we can predict the location of the user $u$ at time $t^u_\star$.

\subsection{Embedding Model}

In order to learn the features of entities from the facts in the observed {\em STKG}, we propose our embedding model in this section. Specifically, this embedding model maps each entity or relation type into a low-dimensional vector, and utilizes a defined scoring function to measure the plausibility of facts in the knowledge graph. Then, the embedding vectors are optimized by maximizing the plausibility of all facts in the observed knowledge graph. Based on the well-learned embedding vectors, we can further implement knowledge graph completion, i.e., predicting the potential relation between entities.

Specifically, for each entity $e\in\mathcal{G}$, we define its embedding vector as $e^E\in\mathbb{C}^d$, where $\mathbb{C}^d$ is the $d$-dimensional complex vector space. It has been shown that compared with real embeddings, complex embeddings have a stronger ability in capturing antisymmetric relations~\cite{Trouillon2016Complex,lacroix2020tensor}. Then, for an arbitrary user $u\in\mathcal{U}$, PoI $p\in\mathcal{P}$, and time $t\in\mathcal{T}$, the scoring function of the tuple $x=(u,r_V,p,t,A)$ is defined as follows:
\begin{equation}\label{equ:score}
f_{r_V}(u,p,t,A)={\rm Re}(<u^E, {r_V^E}(t,A),\bar{p^E}>)=\sum^d_{i=1}u^E_i r_{Vi}^E \bar{p^E_i},
\end{equation}
where $u^E$, $r^E_V$, and $p^E$ are the embedding vector of $u$, $r_V$, and $p$, respectively.
$u^E_i$, $r_{Vi}^E$, and $p^E_i$ are their $i$-th elements, respectively. 
$d$ is the dimensionality of embedding vectors.
In addition, $<\cdot>$ is the multi-linear dot product.  
For an arbitrary complex number $z$, $\bar{z}$ is the complex conjugate of $z$, and ${\rm Re}(z)$ is the real component of $z$.

Note that in (\ref{equ:score}), the embedding vector $r^E_V$ of the spatio-temporal mobility pattern relation is defined as a dynamic function of the time $t$ and auxiliary information $A$. Specifically, the embeddings of entities in $\{t\}\cup A$ are used to modulate $r^E_V$ to obtain an embedding vector dependent on the spatio-temporal context. Formally, the mathematical expression is formulated as follows:
\begin{equation}\label{equ:rvembd}
r_V^E(t,A)=\left[
\begin{array}{l}
r'_V\odot \bar{t^E}\\
r''_V\\
\end{array}
\right] 
\odot A^E,
\end{equation}
where $\odot$ represents the element-wise product. Specifically, (\ref{equ:rvembd}) is the element-wise product of a time-modulated embedding vector and the embedding vector of the auxiliary information $A$. More specifically, we consider that some mobility patterns might be strongly time-dependent while others may be weakly time-dependent. For example, the journey-to-work mobility patterns of a commuter must be strongly time-dependent, while the mobility pattern of a taxi driver might be weakly time-dependent, which might have a stronger correlation with the auxiliary context information, e.g., the previously visited location. Inspired by this phenomenon, we model the first part in (\ref{equ:rvembd}) as two concatenated subvectors, where the first subvector $r'_V\odot \bar{t^E}\in\mathbb{C}^{d_1}$ is explicitly multiplied with the time embedding vector $\bar{t^E}\in\mathbb{C}^{d_1}$, while the second subvector $r''_V$ is not. In addition, $A^E$ is the embedding vector of the auxiliary information $A$. We calculate it by implementing element-wise product of the embedding vectors of all entities in $A$ as follows:
\begin{equation}
A^E = \bigodot\limits_{e\in A}e^E.
\end{equation}
Further, we define the temporal dynamic ratio $\alpha$ as the fraction of the number of dimensions of $r'_V\odot \bar{t^E}$, i.e., $\alpha=d_1/d$.

As for the scoring function of the affiliation relation, we define it in a similar way with (\ref{equ:score}). Specifically, for a PoI $p\in\mathcal{P}$ and a category $c\in\mathcal{C}_i$ with arbitrary $i\in\{1,2,3\}$, the scoring function of the tuple $y=(p,r_{Ci},c)$ is defined as follows:
\begin{equation}\label{equ:CateScore}
f_{r_{Ci}}(p,c)={\rm Re}(<p^E, {r_{Ci}^E},\bar{c^E}>).
\end{equation}

\subsection{Model Optimization}

For each training tuple $x=(u,r_V,p,t,A)$ of the spatio-temporal mobility pattern relation, we use the cross entropy of the visited location as its loss, which can be calculated as:
\begin{equation}\label{equ:visitloss}
\begin{aligned}
\footnotesize
\ell_V(x)  = & \left[ -f_{r_V}(u,p,t,A) +{\rm log} \left(\sum_{p'\in\mathcal{P}}{\rm exp}(f_{r_V}(u,p',t,A))\right)\right].
\end{aligned}
\end{equation}
However, calculating (\ref{equ:visitloss}) requires computing the scoring function of all the $|\mathcal{P}|$ tuples in $\{(u,p',t,A)|p'\in\mathcal{P}\}$, which is time-consuming. Thus, we solve this problem by adopting the strategy of negative sampling. Specifically, for each positive sample $x=(u,r_V,p,t,A)$, multiple negative samples are draw by randomly replacing $p$ by $p'\in\mathcal{P}$. Then, (\ref{equ:visitloss}) is approximated by the following equation:
\begin{equation}\label{equ:visitlossns}
\begin{aligned}
\footnotesize
\ell_V(x)  \approx & \left[ {\rm log}\ \sigma(-f_{r_V}(u,p,t,A)) + \sum^{N_{\rm neg}}_{j=1} E_{p'_j\sim P_n(p'))} \left[{\rm log}\ \sigma(f_{r_V}(u,p',t,A))\right]\right],
\end{aligned}
\end{equation}
where $N_{\rm neg}$ is the number of negative samples, $P_n(p')$ is a noise distribution, and $\sigma$ is the sigmoid function.

\begin{algorithm}[bp!]
\SetAlgoNoLine
    \SetKwBlock{Begin}{Initialize:}{}
    \SetKwBlock{Def}{With probability $p_k/c$ do}{}{}
    \DontPrintSemicolon
  \caption{Training algorithm for {\em STKG}}\label{Alg:train}
  \textbf{Input:} $N_{\rm epoch}$, $N_{\rm neg}$, {\em STKG} $\mathcal{G} = \{\mathcal{G}^V \cup \mathcal{G}^C\}$ as the training set.\\  
  \textbf{Output:} embeddings $\mathcal{U}^E,\mathcal{T}^E,\mathcal{P}^E,\mathcal{C}^E$,$\mathcal{R}^E$.\\
  \textbf{Initialize:} initialise embeddings $\mathcal{U}^E,\mathcal{T}^E,\mathcal{P}^E,\mathcal{C}^E$,${r_V}^E,{r_C}^E$.\\
  \For{i $\in\{1,...,N_{\rm epoch}\}$}
  {    
  $\mathcal{S}\leftarrow \mathcal{G}$\\
  \While{$\mathcal{S}\not=\emptyset$}{
  Sample a mini-batch $\mathcal{S}_{batch} \subset \mathcal{S}$\\
  $\mathcal{S} \leftarrow \mathcal{S}/\mathcal{S}_{batch}$;\\
      $\mathcal{L}\leftarrow 0$\\
      \For{$s \in \mathcal{S}_{batch}$ }
      {   
          \eIf{ s $\in \mathcal{G}^V$}{
          Construct negative sample set $\mathcal{N}_s$;\\
          $f_{r_V}(s) \leftarrow $ compute the score using (\ref{equ:score})\\
          $\ell(s)\ \leftarrow $ compute the loss using (\ref{equ:visitlossns});\\}{  $f_{r_{C_i}}(s) \leftarrow $ compute the score using (\ref{equ:CateScore})\\
          $\ell(s)\ \leftarrow $ compute the loss using (\ref{equ:affloss});\\
          }
          $\mathcal{L} \leftarrow \mathcal{L} + \ell{(s)}$;
          }
        Update parameters of embeddings w.r.t the gradients using $\bigtriangledown \mathcal{L}$;
      }
      }
  \vspace{-0.1cm}
\end{algorithm}

Then, for all spatio-temporal mobility pattern relations contained in $\mathcal{G}^V$, the total loss can be expressed by:
\begin{equation}
\begin{aligned}
\footnotesize
\mathcal{L}_V  = & \sum_{x\in \mathcal{G}^V}\ell_V(x).
\end{aligned}
\end{equation}

Similarly, for each train tuple  $y=(p,r_{Ci},c)$ of the affiliation relation, we use the cross entropy of the affiliated PoI category as the loss, which can be expressed as:
\begin{equation}\label{equ:affloss}
\begin{aligned}
\footnotesize
\ell_{Ci}(y)  = & \left[ -f_{Ci}(p,c) +{\rm log} \left(\sum_{c'\in\mathcal{C}_i}{\rm exp}(f_{Ci}(p,c'))\right)\right].
\end{aligned}
\end{equation}
Since the number of PoI categories is much less compared with the number of PoIs, we do not implement negative sampling in calculating (\ref{equ:affloss}). Then, the total loss of all affiliation relation contained in $\mathcal{G}^{Ci}$ can be calculated by:
\begin{equation}
\begin{aligned}
\footnotesize
\mathcal{L}_{Ci}  = & \sum_{y\in \mathcal{G}^{Ci}}\ell_{Ci}(y).
\end{aligned}
\end{equation}

Based on the above formulation of the loss in terms of different types of relations, the total loss of the {\em STKG} $\mathcal{G}$ can be expressed as follows:
\begin{equation}\label{equ:totalloss}
\begin{aligned}
\footnotesize
\mathcal{L}  = \mathcal{L}_V  + \beta\sum^3_{i=1}\mathcal{L}_{Ci},
\end{aligned}
\end{equation}
where $\beta>0$ is the parameter to balance the influence of the loss of different types of relations.
Overall, we can train our model by minimizing the total loss~(\ref{equ:totalloss}), and obtain the embedding vectors of all entities in $\mathcal{G}$.

\begin{algorithm}[b!]
\SetAlgoNoLine
    \DontPrintSemicolon
  \caption{Predicting future movement based on {\em STKG}.}\label{Alg:pred}
  \textbf{Input:} The set of target time $\{t^{u}_\star\}_{u\in\mathcal{U}}$, the corresponding auxiliary information $\{A^u_\star\}_{u\in\mathcal{U}}$, and the embeddings of all entities including  $\mathcal{U}^E,\mathcal{T}^E,\mathcal{P}^E,\mathcal{C}^E$,$\mathcal{R}^E$.\\
  \textbf{Output:} $\{p^u_\star\}_{u\in\mathcal{U}}$, where each $p^u_\star$ presents the PoI predicted to be visited by $u$ at time $t^{u}_\star$.\\
  \For{$u\in\mathcal{U}$}
  {
      \For{$p\in\mathcal{P}$}
      {
       $f_p$ = $f_{r_V}(u,p,t^u_\star,A^u_\star)$ using (1)\\
      }
      $p^u_\star = \mathop{\arg\max}\limits_{p}f_p$\\
  }
\end{algorithm}

\subsection{Algorithm Analysis}
We present the pseudocode describing the process of learning embedding vectors of entities in Algorithm~\ref{Alg:train}. As we can observe, in each epoch, this algorithm samples a mini-batch of facts in the training set $\mathcal{S}$. Then, for each factor in the mini-batch, its loss is calculated based on (\ref{equ:visitlossns}) and (\ref{equ:affloss}). Finally, this algorithm updates embeddings of entities based on the gradient of the total loss. Similarly, the pseudocode describing the process of implementing mobility prediction based on the proposed {\em STKG} is shown in Algorithm~\ref{Alg:pred}. As we can observe, for each user $u$ and the corresponding target time $t^u_\star$ and auxiliary information $A^u_\star$, this algorithm iterates over all candidate PoI $p$ in $\mathcal{P}$ and compute the score $f_{r_V}(u,p,t^u_\star,A^u_\star)$. Then, the PoI with the largest score is found as the prediction results. In the practical scenarios, according to the requirements of mobility prediction tasks, this algorithm can also rank candidate PoIs based on their scores and reserve top-$k$ PoIs as the prediction results.

\begin{figure*}[t]
\centering
\subfigure{\includegraphics[width=.6\textwidth]{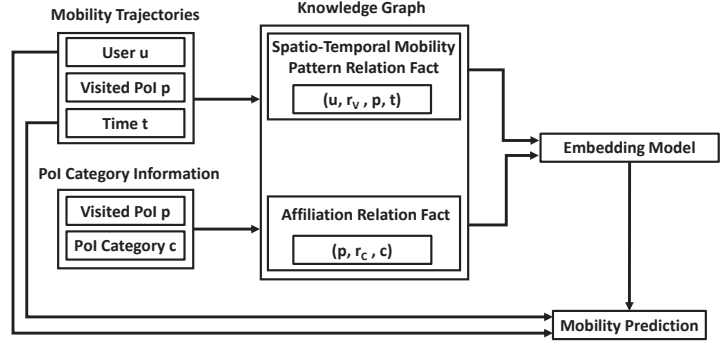}}
\caption{An example process of our proposed method.} \label{fig:actual_example}
\end{figure*}

We have further utilized a simple example to illustrate the process of our method in Figure~\ref{fig:actual_example}. Specifically, we first extract user, PoI, time from mobility trajectories and PoI categories. Then, the facts of the spatial-temporal mobility pattern relation and affiliation relation are extracted to construct our knowledge graph. Through the embedding model, we can get the learned embeddings of the user, PoI, and time. Finally, we can compute the scores of all PoIs given user, time information, and the top-$k$ PoIs are our prediction results based on the scores. 

\para{Complexity.} As we can observe from Algorithm~\ref{Alg:train}, the update of each fact in $\mathcal{G}^V$ and $\mathcal{G}^C$ cost $\mathcal{O}(d N_{\rm neg})$ time and $\mathcal{O}(d |\mathcal{C}|)$ time, respectively, where $d$ is the dimensionality of embedding vectors. Thus, the total computational complexity of the training algorithm is $\mathcal{O}(N_{\rm epoch}d |\mathcal{G}|(|\mathcal{C}|+N_{\rm neg}))$. Differently, in the prediction algorithm, the negative sampling strategy cannot be used. Thus, its total computational complexity is $\mathcal{O}(d |\mathcal{P}||\mathcal{U}|)$. Overall, the computational complexity grows linearly with the problem size, which is feasible in practice.

\section{Evaluation}\label{sec:evaluation}
\subsection{Experimental Settings}

\subsubsection{Datasets}

\begin{figure*}[t!]
\centering
\subfigure[Probability distribution function of time intervals of adjacent mobility records]{\label{Fig:pdf}
\includegraphics[width=.40\textwidth]{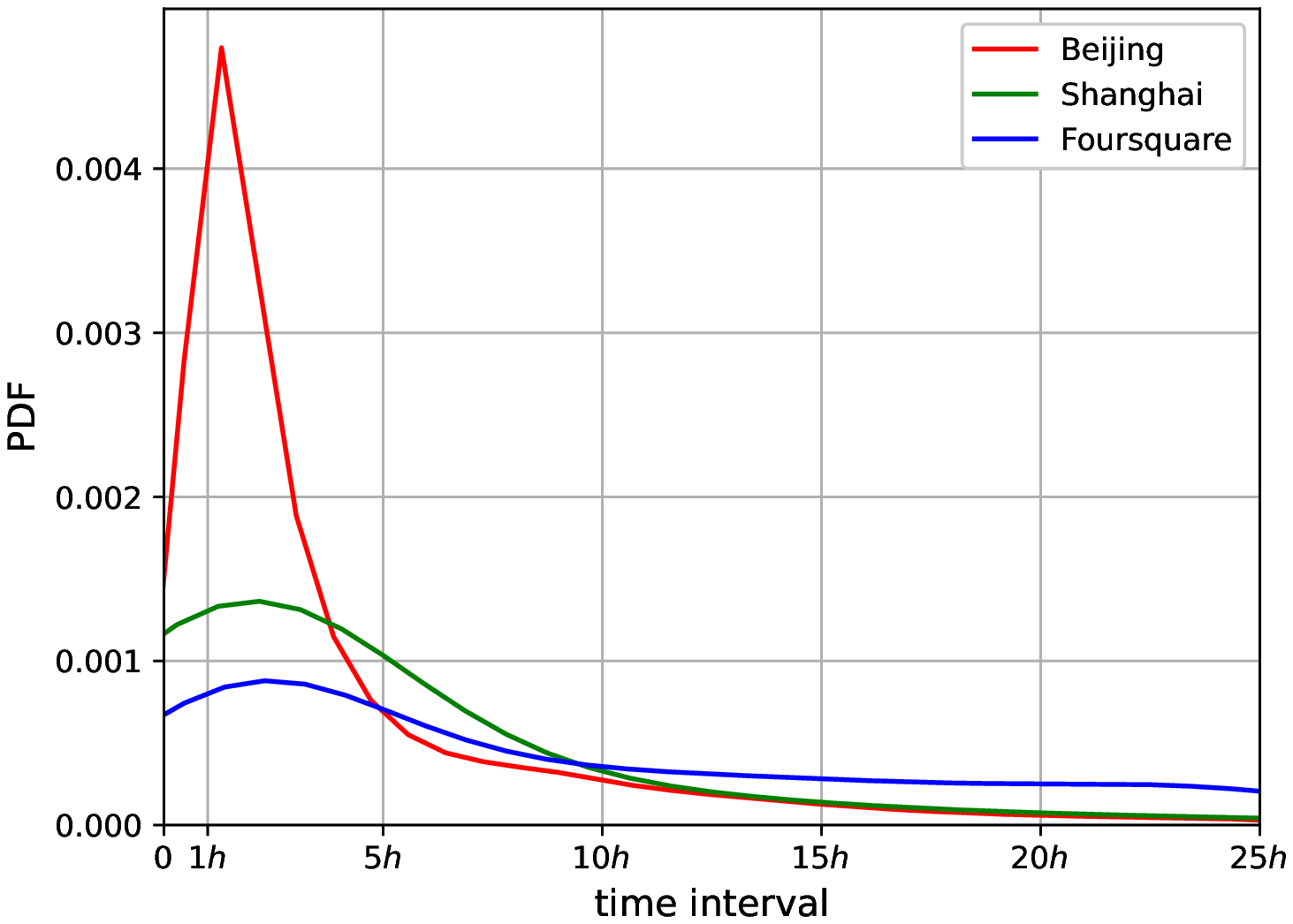}}
\hspace{0.2cm}
\subfigure[Cumulative distribution function of time intervals of adjacent mobility records]{\label{Fig:cdf}
\includegraphics[width=.40\textwidth]{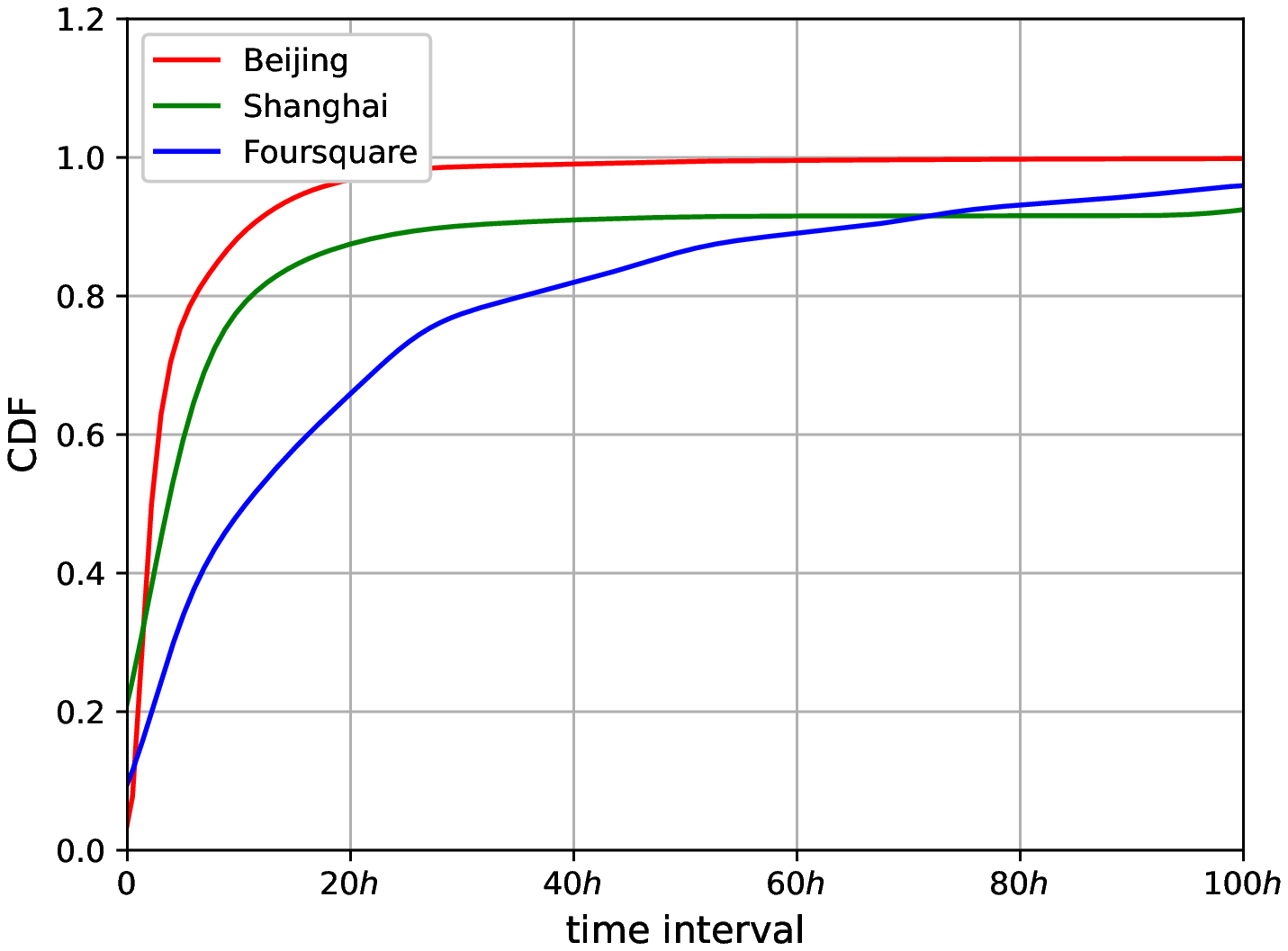}}
\caption{Detailed information of three mobility datasets.}\label{fig:dataset}
\end{figure*}

We have utilized three representative datasets to evaluate the performance of our experiment. The first one is a publicly available check-in data of New York from a previous work~\cite{yang2014modeling}, which is collected from Foursquare with the duration from April 12, 2012 to February 16, 2013. Each record in this dataset includes the corresponding user ID, timestamp, and PoI. The other two real-world human mobility datasets are collected from WeChat, which is one of the most popular mobile social networks in China. Specifically, users' locations are recorded in the form of GPS coordinates when they access location services provided by WeChat through their applications, e.g., check-in, sharing locations, etc. Further, the GPS coordinates will be mapped into different PoIs if they fall within the coverage area of the corresponding PoIs, which is obtained based on the APIs provided by WeChat.
The first WeChat dataset records users' trajectories from September 17 to October 31 in 2016, covering the whole urban area in Beijing.
The latter WeChat dataset consists of the locations where users have visited in Shanghai from August 17 to November 8, 2018.
Each record in these two datasets includes the user ID, timestamp, and the corresponding PoI.
To better understand the characteristics of the three datasets, we analyze the detailed information of mobility data and the distribution of the time intervals between adjacent mobility records. From Figure~\ref{Fig:pdf}, we can observe that the time intervals of adjacent records on the Beijing dataset are much shorter than the Shanghai dataset and the Foursquare dataset, and 70$\%$ of the time intervals are shorter than 4 hours. Besides, we find that the time intervals in these three datasets follow a similar long-tail distribution.

	\begin{table}
    \centering
	\begin{tabular}{| l | l | l | l |}
		\hline			
		\textbf{Dataset} & \textbf{Beijing} & \textbf{Shanghai}& \textbf{Foursquare}\space\space\space\space\space\space \\ \hline  \hline
		\# Users & 3,083 & 926 & 1,083 \\ \hline
		\# PoIs & 12,597 & 22,627 & 38,333\\ \hline
		\# Records & 650,578 & 114,811 & 227,428 \\ \hline
		\# Fine-level categories & 204 & 281 & 251\\ \hline
		\# Mid-level categories & 49 & 55 & - \\ \hline
		\# Coarse-level category & 14 & 14 & - \\ \hline
	    \# Duration & 45 days & 82 days & 11 months\\ \hline
	\end{tabular}
	\caption{Statistics of three datasets.}
	\label{tab:info_data}
	\vspace*{-5mm}
	\end{table}

\begin{figure}[t]
\centering
\subfigure[Trajectory example of users]{\label{tab:example_v}
\begin{minipage}{0.8\textwidth}
\centering
{\renewcommand\arraystretch{.3}
\begin{tabular*}{6cm}{c}
\tiny
\\
\\
\\
\\
\end{tabular*}}
\footnotesize
\begin{tabular}{|c|c|c|c|c|c|c|c|c|c|c|c|c|c|c|c|c|}
\hline
~         & \multicolumn{16}{c|}{Trajectories} \\ \hline
{User} & {$u_1$} & {$u_1$} & {$u_1$} & {$u_1$} & {$u_1$} & {$u_1$} & {$u_1$} & {$u_1$} & {$u_1$} & {$u_1$} & {$u_2$} & {$u_2$} & {$u_2$} & {$u_2$} & {$u_2$} & {$u_2$} \\
\hline
Time & $t_1$ & $t_2$ & $t_3$ & $t_4$ & $t_5$ & $t_6$ & $t_7$ & $t_8$ & $t_9$ & $t_{10}$ & $t_1$ & $t_2$ & $t_3$ & $t_4$ & $t_5$  & $t_6$      \\\hline
PoI &  $p_A$  & $p_B$  & $p_B$  & $p_C$  & $p_A$  & $p_B$  & $p_A$  & $p_C$  & $p_B$ & $p_A$  & $p_C$ & $p_B$ & $p_A$ & $p_B$ & $p_A$ & $p_C$\\\hline
\end{tabular}
\renewcommand\arraystretch{.3}
\begin{tabular*}{5cm}{c}
\tiny
\\
\\
\\
\end{tabular*}
\end{minipage}
}

\subfigure[Three-level category example of PoIs]{\label{tab:example_c}
\begin{minipage}{0.45\textwidth}
\centering
{\renewcommand\arraystretch{.3}
\begin{tabular*}{8.5cm}{c}
\tiny
\\
\\
\\
\\
\end{tabular*}}
\footnotesize
\begin{tabular}{|c|c|c|c|}
\hline
{Location} & {Fine-level} & {Mid-level}  & {Coarse-level} \\
\hline
$p_A$ & $c_{A1}$ & $c_{A2}$& $c_{A3}$        \\\hline
$p_B$ &  $c_{B1}$  & $c_{B2}$ & $c_{B3}$      \\\hline
$p_C$ &  $c_{C1}$  & $c_{C2}$ & $c_{C3}$      \\\hline
\end{tabular}
\renewcommand\arraystretch{.3}
\begin{tabular*}{8cm}{c}
\tiny
\\
\\
\\
\end{tabular*}
\end{minipage}
}
\subfigure[Example for the mobility prediction process]{\label{fig:example_p}
\begin{minipage}{0.45\textwidth}
\centering
\includegraphics[width=.9\textwidth]{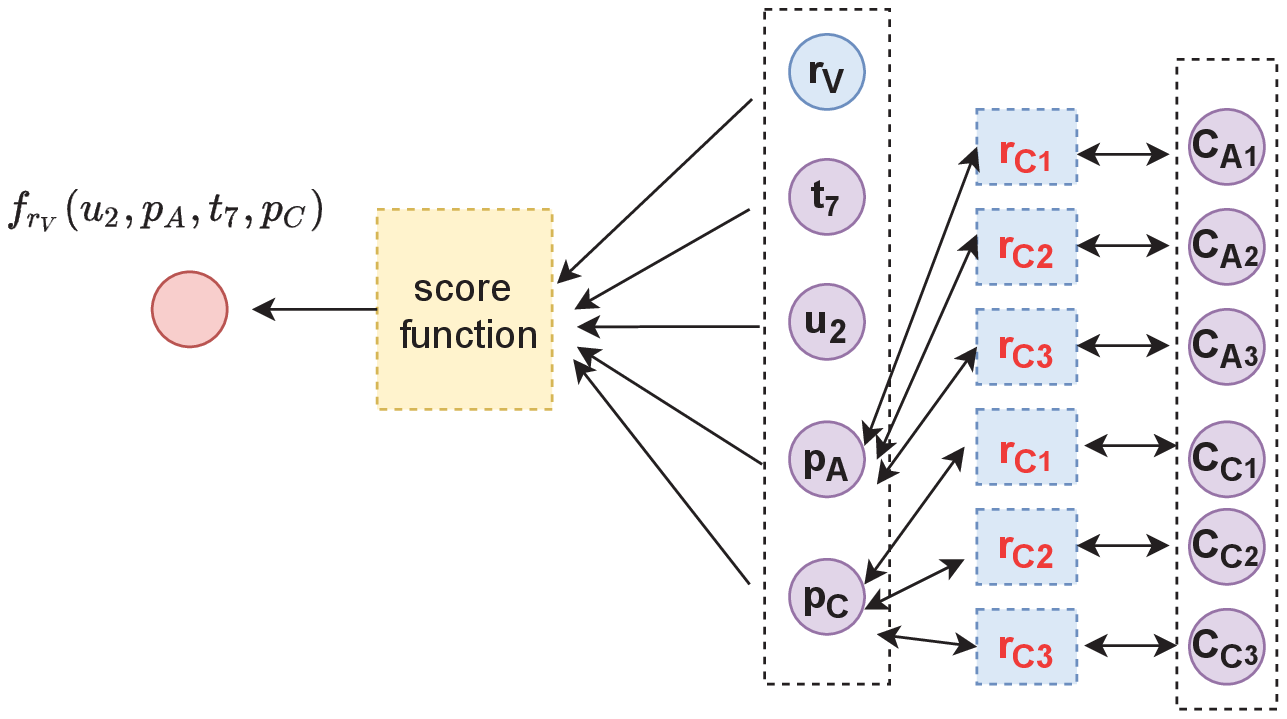}
\end{minipage}
}

\caption{Trajectory and PoI category example of data format.}\label{fig:example}
\end{figure}

To help extract the information of PoIs and increase the accuracy of prediction, we have also utilized three levels of PoI categories. In the Beijing dataset, the three levels of categories consist of 204 fine-level categories, 49 mid-level categories, and 14 coarse-level categories, respectively. As the number of categories increases, the representation of the PoI category is more fine-grained. For example, the `restaurant' of coarse-level categories contains eight mid-level categories, such as `Chinese food', `Southeast Asian cuisine', etc. Further, `Beijing food' and `Sichuan food' in fine-level categories are affiliated to `Chinese food' in mid-level categories.  As similar to the Beijing dataset, the PoIs of the Shanghai dataset also have three levels of categories, including 281 fine-level categories, 55 mid-level categories, and 14 coarse-level categories. Different from them, the Foursquare dataset only contains 251 fine-level categories.
To illustrate the details of the data format clearly, we have listed an example about trajectories and PoIs with their categories in Figure~\ref{fig:example}, respectively. For simplicity, we only leverage three PoIs $p_A$, $p_B$, and $p_C$ in the trajectory example. Figure~\ref{tab:example_v} presents the mobility records of users $u_1$ and $u_2$, e.g., the second column represents user $u_1$ visited the PoI $p_A$ at the timestamp $t_1$, and the first 10 records constitute the trajectory of user $u_1$ and the left 6 records constitute the trajectory of user $u_2$. Figure~\ref{tab:example_c} demonstrates the three-level category structure of the PoIs.
Finally, we present a diagram of utilizing the relational facts to extract the embeddings of entities, and finally combine the obtained embeddings with the scoring function to implement the mobility prediction in Figure~\ref{fig:example_p}.

In this paper, we have implemented a number of pre-processing steps on the three datasets. First, we discretize the timestamps of the three datasets into 30 minutes.
Further, since the mobility records from WeChat are collected passively, the users with fewer records rarely go outside and their mobility patterns exhibit strong regularity, leading to an extremely high accuracy of predicting their future movement and affecting the training process of our model.
Therefore, we have filtered out the users with less than 30 records and visited less than 5 places. In general, through the mechanism of filtering, we delete the 15\%, 13\%, and 5\% users of the original data on the Beijing, Shanghai, and Foursquare datasets, respectively.

The datasets are split by 7:1:2 as the train-valid-test datasets. Specifically, the mobility records of each user are first ranked according to their timestamps, and then the first 70$\%$ records of each user's trajectory are regard as the training set, the middle 10$\%$ records are selected as the valid dataset, and the last 20$\%$ records are used as the test dataset to evaluate the performance of our model.
The basic information of the three datasets is presented in Table~\ref{tab:info_data}, including the cities,  number of users, number of locations, duration, number of three-level categories and the whole number of records.
Taking Beijing as an example, Table~\ref{tab:categories} shows a part of the involved PoI categories with different levels.

    \begin{table}
    \begin{center}
    \small
    \begin{tabular}{|c|l|l|}
        \hline
        ID & Coarse-level Category & Mid-level Category \\  \hline
        \#1 & Food & Chinese food, Western food, Dessert. \\ \hline
        \#2 & Shopping & Market, Mall, Bazaar.  \\ \hline
        \#3 & Leisure sports & KTV, Cinema, Gym, Concert.  \\ \hline
        \#4 & Accommodation & Tavern, Hotel, Homestay.  \\ \hline
        \#5 & Business & Finance, Company, Firm, Corporation.  \\ \hline
        \#6 & Residence  & Office building, Residential district.  \\ \hline
        \#7 & Life services & Store, Supermarket, Laundry, Home service.  \\ \hline
        \#8 & Transportation & Bus station, Dock, Subway station, Service area. \\ \hline
        \#9 & Car services & Gas station, Parking lot.  \\ \hline
        \#10 & Education & Campus, School, Vocational college.  \\ \hline
        \#11 & Medical services & Hospital, Clinic, First-aid center.  \\ \hline
        \#12 & Holiday & Museum, Beach, Park, Art gallery, Zoo.  \\ \hline
        \#13 & Organizations  & Government organizations, Scientific institution.  \\ \hline
        \#14 & Factory & Agriculture factory, Enterprise.  \\ \hline

    \end{tabular}

   \caption{The utilized PoI categories and taxonomies.}
   \label{tab:categories}
   \vspace*{-5mm}
    \end{center}
    \end{table}

Considering the mobility data containing much private information, we have taken active actions to protect the privacy of the involved users. First, the users' IDs in our experiment are all anonymized, which do not contain any personally identifiable information. Besides, all the researchers are regulated by a strict non-disclosure agreement to prevent the leakage of important information. Last, our work has received approval from the provider of the datasets.

\subsubsection{Compared Algorithms}

We elaborately select the following mobility prediction algorithms to be compared with our proposed model,
which covers the most representative mobility prediction methods proposed in recent years.

\begin{itemize}
\item DeepMove~\cite{Feng2020Predicting}: This method incorporates the attention mechanism in the recurrent neural network for better modeling long-term correlation of the future movement with the historical mobility records.
\item ROI~\cite{Jiang2018Deep}: This method is based on recurrent neural networks and incorporates the small number of important regions of users, which are denoted as the region of interest (ROI).
\item APHMP~\cite{fan2019decentralized}: The method is another mobility prediction model based on the recurrent neural network combined with the attention mechanism.
\item LSTM~\cite{abu2018will}: This model utilizes the Long Short Term Memory (LSTM) network to model the sequence of mobility records and implement mobility prediction.
\item ARNN~\cite{GuoSunZhangTheng2020}: This method utilizes the semantic and spatial information obtained from the construct knowledge graph and makes mobility predictions based on the combination of the LSTM network and the attention mechanism.
\end{itemize}

In our comparative experiments, we leverage the released source code of DeepMove\footnote[1]{https://github.com/vonfeng/DeepMove}, while other methods are all re-implemented by us. Our final parameter settings have a little difference from the original study due to the different characteristics of the datasets used in our experiments.
Specifically, for the DeepMove algorithm, the learning rate is set as 0.0005 and we select the sequential encode attention module and average sampling strategy to conduct the experiments.
For the LSTM algorithm, the learning rate and dropout are set as 0.001 and 0.5, respectively. For the ROI model, we evaluate the performance regarding the defined Regions-of-Interest (ROIs) as PoIs in our experiment. In the APHMP model, we delete the module of decentralized learning since the privacy issues are not in the scope of our paper.
In addition, the knowledge graph in ARNN is constructed based on the semantic information by utilizing the PoI category information in our datasets.

There are also a number of probability-based mobility prediction algorithms including HMM~\cite{Mathew2012Predicting}, PMM~\cite{cho2011friendship}, GMove~\cite{zhang2016gmove}, etc. However, deep learning based models have been shown to have a higher performance than the probability-based models~\cite{Feng2020Predicting,fan2019decentralized}. In addition, among the deep learning based models, DeepMove has been the most representative algorithm with one of the best performance. Thus, we only compare our proposed algorithm with the selected five baseline algorithms.

\subsubsection{Evaluation Metrics}
In our experiment, four metrics are utilized to evaluate the performance of mobility prediction based on our model. First is the mean reciprocal rank (MRR) metric, which is widely used in knowledge graphs (KGs). The other three metrics are Acc@1, Acc@5 and Acc@10, representing the correctness of the top-1, top-5 and top-10 predicted locations, which are formally defined as follows.
\begin{equation}\label{equ:metrics}
\begin{aligned}
\footnotesize
   MRR = \frac{1}{N}& \sum_{i=1}^N \frac{1}{rank_i}.  \hspace{10mm}
Acc@k = \frac{1}{N}\sum_{i=1}^N {rank_i \leq k}.
\end{aligned}
\end{equation}
Specifically, $N$ represents the number of mobility records used in the test set.
In addition, $rank_i$ represents the rank of $i_{th}$ record. In our experiment, we first compute the metrics
(\ref{equ:metrics}) for each user, and then compute their average value as the final indicator of the mobility prediction performance.
Specifically, MRR is the average of the reciprocal of the ranks of the predicting results, while Acc@k measures the correctness of the top-k candidates.
They all range from 0 to 1, where a higher MRR or Acc@k indicates a better prediction result.

\subsubsection{Parameter Setting}
In our experiment, the default embedding size of the user, location, categories, and the two types of relations are all set to be 100. The hyper-parameters in our models include the learning rate, embedding regularization coefficient, time regularization coefficient, temporal dynamic ratio $\alpha$ and parameter $\beta$. On all datasets, the embedding regularization coefficient and time regularization coefficient are set to be 0.01, while the learning rate ranges from 0.01 to 0.1. Considering the time-dependent nature of the mobility pattern, the default value of temporal dynamic ratio $\alpha$ is 0.5, and with the increase of the value of $\alpha$, we could observe the dynamic change of performance. Because of the difference in data quality, we have utilized different values of $\beta$ to balance the influence of the loss of different types of relations, i.e., 20 in the Beijing dataset, 5 in the Shanghai dataset, and 10 in the Foursquare dataset.

\subsection{Experimental Results}
\subsubsection{Overall Performance} In Table~\ref{tab:result} and Table~\ref{tab:foursquare}, we have evaluated the performance of our proposed model and baselines on the three datasets. From the statistical results, we can observe that our model achieves the best performance in terms of the four metrics, indicating the superiority of knowledge graph embedding techniques on the problem of human mobility prediction. By comparing the performance of our proposed model on the three datasets, we can observe that the results on the Beijing dataset are better than those of the Shanghai dataset and Foursquare dataset. That's because users in the Beijing dataset have denser trajectories, and the mobility records are collected with a higher frequency than the Shanghai dataset, while the Foursquare check-in dataset holds the sparest records for each PoI and only has the fine-level categories, which leads to the disability of capturing two types of affiliation relationships than the other two datasets.
Further, we could observe that our proposed method achieves obvious performance gain compared with ARNN on all three datasets.
Actually, ARNN feeds the embeddings of PoIs and PoI categories from the constructed knowledge graph into the LSTM directly, neglecting the dynamic changes within the training period.
Thus, the prediction accuracy of our model is about 8.91\% better than ARNN in terms of Acc@1 on the Beijing dataset, suggesting
that the knowledge graph technique is a more effective model to utilize the spatio-temporal context information compared with directly feeding them into the RNN models.
Besides, apart from ARNN, DeepMove also achieves a more accurate prediction on the Beijing dataset than other baselines, demonstrating the effectiveness of the attention mechanisms of historical trajectories. In summary, our proposed model has a stronger generalization ability and achieves more accurate human mobility prediction.

\begin{table*}
        \centering
		\begin{tabular}{|l||l|l|l|l||l|l|l|l|}
			\hline
			~         & \multicolumn{4}{c||}{Beijing} & \multicolumn{4}{c|}{Shanghai} \\ \hline
			~         & \multicolumn{3}{c|}{Accuracy@k} & \multirow{2}{*}{MRR}  & \multicolumn{3}{c|}{Accuracy@k} & \multirow{2}{*}{MRR}  \\ \cline{1-4}\cline{6-8}
			~         & k=1        & k=5    & k=10    &                     & k=1        & k=5    & k=10    &                     \\ \hline
			LSTM   & 38.17\% & 52.93\% & 55.56\% & 44.96\%               & 34.01\% & 50.04\% & 52.36\% & 40.66\%               \\ \hline
			ROI   & 35.25\% & 47.73\% & 50.25\% & 41.10\%               & 32.12\% & 46.46\% & 48.77\% & 38.47\%               \\ \hline			
			APHMP   & 40.46\% & 57.28\% & 60.07\% & 48.14\%               & 35.03\%    & 53.02\% & 55.31\% & 43.18\%             \\ \hline
			DeepMove   & 41.04\% & 57.55\% & 60.22\% & 48.61\%              & 35.74\%    & 55.15\% & 57.42\% & 44.51\%             \\ \hline
			ARNN   & \underline{44.18\%} & \underline{62.93\%} & \underline{68.47\%} & \underline{53.19\%}               & \underline{37.87\%}     &  \underline{58.03\%}   & \underline{60.48\%}  & \underline{46.92\%}          \\ \hline
			Our   & \textbf{53.09\%} & \textbf{76.75\% }& \textbf{79.97\%} & \textbf{63.66\%}             &
			\textbf{42.91}\%     & \textbf{66.10}\% & \textbf{69.62}\% & \textbf{52.75}\%                \\ \hline
		\end{tabular}
		\caption{Performance on two datasets, where bold denotes best (highest) results and underline denotes the second best results.}
	    \label{tab:result}
	\end{table*}

	\begin{table*}
        \centering
		\begin{tabular}{|l||l|l|l|l|}
			\hline
			~         & \multicolumn{4}{c|}{Foursquare} \\ \hline
			~         & \multicolumn{3}{c|}{Accuracy@k} & \multirow{2}{*}{MRR}    \\ \cline{1-4}
			~         & k=1        & k=5    & k=10    &                                    \\ \hline
			LSTM   & 8.20\% & 17.06\% & 20.03\% & 12.33\%                 \\ \hline
		    ROI   & 6.69\% & 14.49\% & 17.33\% & 10.35\%       \\ \hline			
			APHMP   & 8.29\% & 18.21\% & 22.05\% & 12.88\%      \\ \hline
			DeepMove   & 10.13\% & 19.92\% & 23.45\% & 14.73\%   \\ \hline
			ARNN   & \underline{12.59\%} & \underline{23.37\%}     & \underline{26.43\%}     &  \underline{17.21\%}     \\ \hline
			Our   & \textbf{17.71\%} & \textbf{32.35\%} & \textbf{37.22\%} & \textbf{24.34\%}             \\ \hline
		\end{tabular}
		\caption{Performance on the Foursquare dataset, where bold denotes best (highest) results and underline denotes the second best results.}
	    \label{tab:foursquare}
	\end{table*}

\subsubsection{Computational Time}
\begin{figure}[t]
\centering
\subfigure[Training time vs. $|\mathcal{G}|$ ]{\label{Fig:Train}
\includegraphics[width=.32\textwidth]{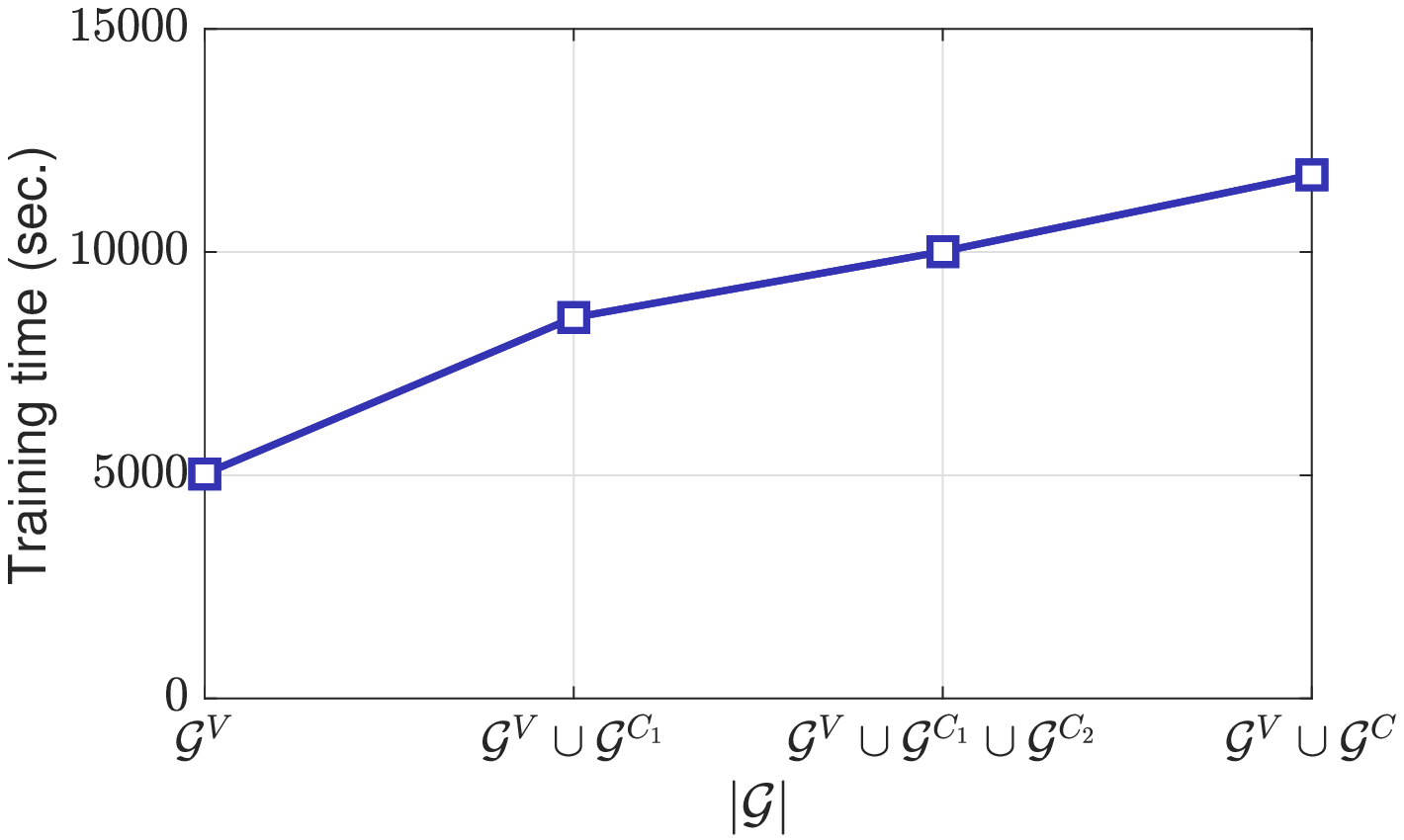}}
\subfigure[Test time vs. $|\mathcal{U}|$ ]{\label{Fig:Test_user}
\includegraphics[width=.32\textwidth]{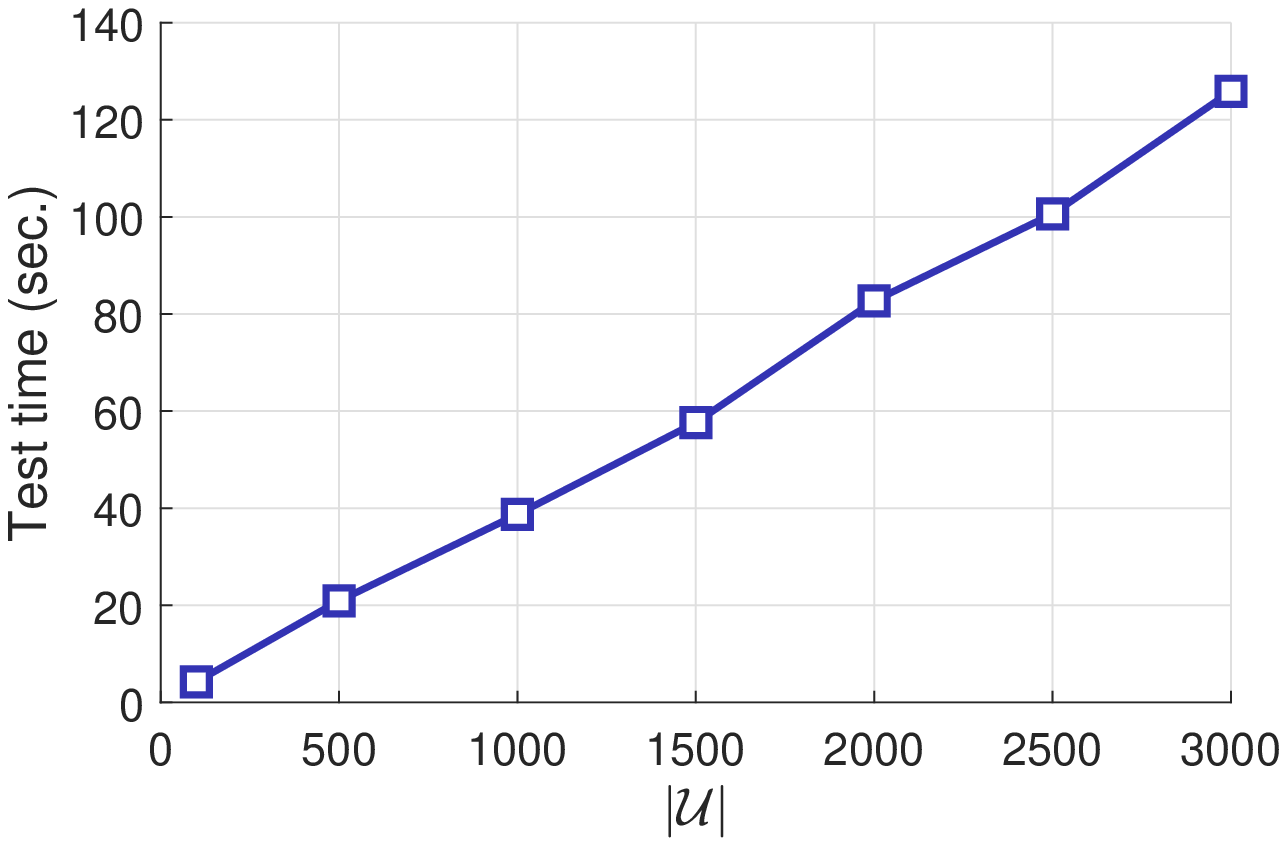}}
\subfigure[Test time vs. $|\mathcal{P}|$]{\label{Fig:Test_poi}
\includegraphics[width=.32\textwidth]{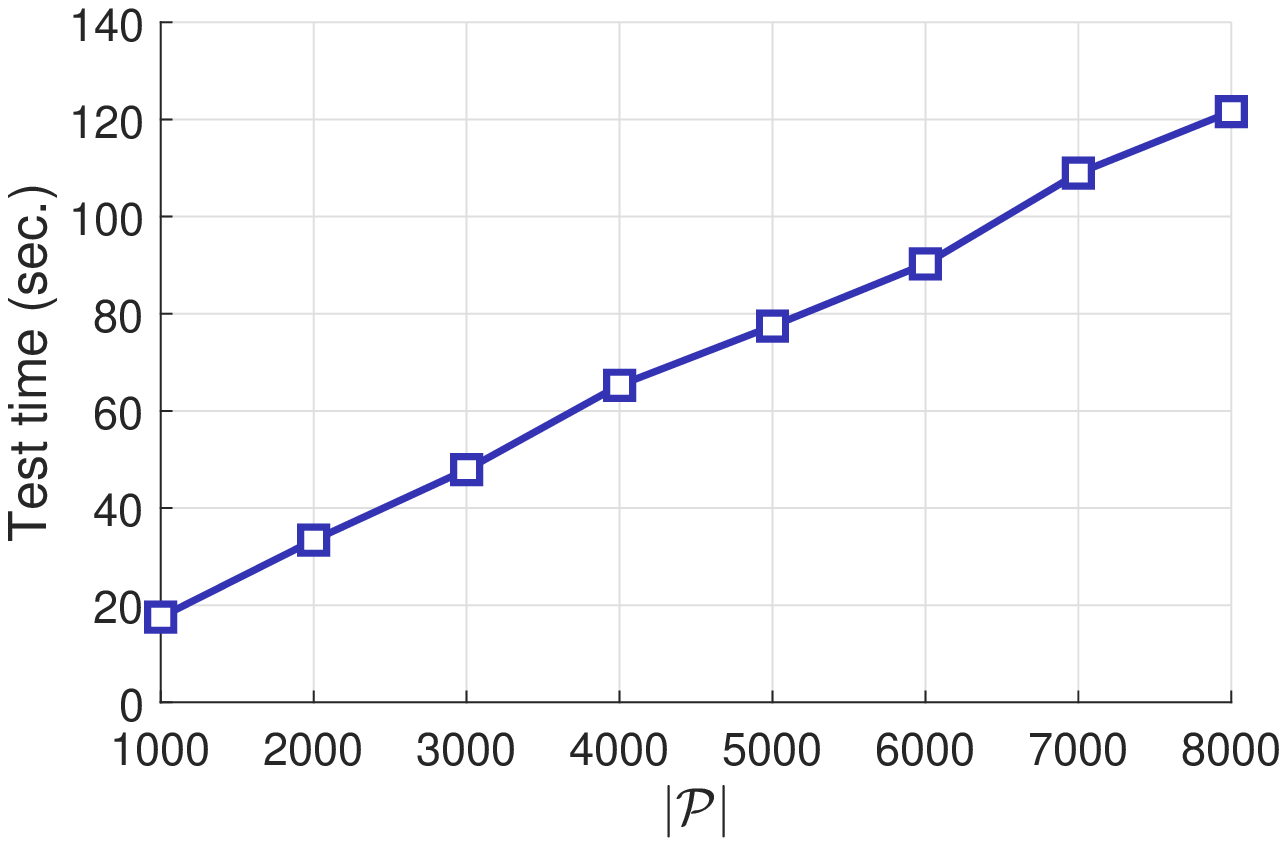}}
\caption{Details of the computational complexity (a) training computational time vs. the number of facts in the {\em STKG}, i.e., $|\mathcal{G}|$, (b) test computational time vs. the number of users $|\mathcal{U}|$, and (c) test computational time vs. the number of PoIs $|\mathcal{P}|$ on the Beijing dataset.} \label{fig:complexity}
\end{figure}

To illustrate the computation complexity of our model, we evaluate the performance of our model in terms of the training time and test time.
Figure~\ref{fig:complexity} shows the corresponding results. Specifically, all the experiments are conducted on a 48-Core 2.20GHz Linux server with 8 Titan Xp GPU. The experiment of training time is implemented with the different number of PoI categories, and the average training time is shown in Figure~\ref{Fig:Train}, while the average test time is shown as the functions of the number of users and PoIs in Figure~\ref{Fig:Test_user} and Figure~\ref{Fig:Test_poi}, respectively.
In Figure~\ref{Fig:Train}, we can observe that the training time exhibits a linear growth with the number of facts in our {\em STKG}, i.e., $|\mathcal{G}|$, which conforms to the total computational complexity of the training algorithm. From Figure~\ref{Fig:Test_user} and Figure~\ref{Fig:Test_poi}, we can also observe that the test time has a linear relation with $|\mathcal{U}|$ and $|\mathcal{P}|$. That is, the corresponding complexity grows with the product of the number of PoIs and the number of users.

Besides, we illustrate the experimental results on the three datasets to compare different methods in terms of the total computational time in Table~\ref{tab:total_time}. Due to the quadratic computational complexity of the attention mechanism, the models of APHMP, DeepMove, and ARNN all have about a quadratic growth of the time complexity with the length of the historical trajectories, while LSTM, ROI, and our model do not have this issue and only require to iterate over each record without the computational time influenced by the historical trajectories, indicating their strong ability in terms of scalability. From the total computational time shown in Table~\ref{tab:total_time}, we can observe that our proposed method costs the least time compared with other methods.

	\begin{table}
    \centering
	\begin{tabular}{| l | l | l | l |}
		\hline			
		\textbf{ } & Beijing(sec.) & Shanghai(sec.) & Foursquare(sec.)\space\space\space\space\space \\ \hline  \hline
		LSTM & 11,503 & 2,464 & 9,784\\ \hline
		ROI & 9,000 & 2,062 & 9,628\\ \hline
		APHMP & 12,946 & 3,041 & 11,601\\ \hline
		DeepMove & 13,129 & 3,796 & 12,190 \\ \hline
		ARNN & 13,921 & 5,219 & 13,049\\ \hline
		Our & 6,542 & 1,494 & 4,131 \\ \hline
	\end{tabular}
	\caption{The total computational time of three datasets.}
	\label{tab:total_time}
	\vspace*{-5mm}
	\end{table}
	
\subsubsection{Impact of different auxiliary information} In Table~\ref{tab:result_aux}, by applying different types of spatio-temporal mobility pattern relations to predict the PoI the user will go to, we can obtain several observations. First, compared with the basic {\em STMPR}, the model considering the user's last visited PoI achieves better performance in terms of Acc@$1$, Acc@$5$, and MRR. Simply speaking, users prefer to go through similar trajectories. For example, compared with the relation $r_{V_0}$, relation $r_{V_1}$ on $\mathcal{G}^{V}$ improves the mobility prediction performance in terms of MRR by 1.06$\%$ on the Beijing dataset and 2.89$\%$ on the Shanghai dataset, demonstrating the importance of considering the last visited venues. Second, considering the categories of the last visited PoI is much better than considering PoI solely because the PoI category will share information with the affiliation relation between PoIs and PoI categories, thus obtaining more promising results on mobility prediction. Besides, with the generalization of representation on the last visited spatial venue, the model can capture more helpful information on predicting the next venue. From the Table~\ref{tab:result_aux}, we can observe that the model utilizing $r_{V_4}$ has a similar result with  $r_{V_3}$, while increasing the performance by 2.14$\%$ in terms of MRR than $r_{V_1}$ on the Beijing dataset. Furthermore, we also find that the model with information about the categories of PoI is much better than other models, which shows the importance and effectiveness of the auxiliary information. Generally speaking, by leveraging the auxiliary information and applying this information to our proposed model, our proposed methods achieve significant performance gain, demonstrating the effectiveness of the spatio-temporal context information and auxiliary information.

	\begin{table*}
        \centering
		\begin{tabular}{|l|l||l|l|l|l||l|l|l|l|}
			\hline
			\multirow{3}{*}{\em STMPR}    & \multirow{3}{*}{\em STKG}     & \multicolumn{4}{c||}{Beijing} & \multicolumn{4}{c|}{Shanghai} \\ \cline{3-10}
			~    & ~     & \multicolumn{3}{c|}{Accuracy@k} & \multirow{2}{*}{MRR}  & \multicolumn{3}{c|}{Accuracy@k} & \multirow{2}{*}{MRR}  \\ \cline{3-5}\cline{7-9}
			~    & ~     & k=1        & k=5    & k=10    &                     & k=1        & k=5    & k=10    &                     \\ \hline
			$r_{V_0}$ & $\mathcal{G}^{V}$   & 49.85\% & 74.67\% & 77.57\% & 60.09\%               & 36.57\%   & 64.73\% &\ 68.21\% & 48.15\%               \\ \hline
			$r_{V_1}$ & $\mathcal{G}^{V}$   & 50.81\% & 74.55\% & 78.18\% & 61.15\%               & 39.06\%     & 64.90\% & 67.93\% & 51.04\%               \\ \hline
			$r_{V_2}$ & $\mathcal{G}^{V}$   & 51.57\% & 75.81\% & 79.56\% & 62.39\%               & 41.12\%     & 64.27\% & 67.95\% & 51.12\%               \\ \hline
			$r_{V_3}$ & $\mathcal{G}^{V}$   & 52.48\% & 76.16\% & 79.73\% & 63.08\%               & 41.32\%     & 65.07\% & 68.80\% & 51.53\%               \\ \hline
			$r_{V_4}$ & $\mathcal{G}^{V}$   & 52.56\% & \underline{76.54\%} & \textbf{80.05}\% & 63.29\%               & 41.41\%     & 65.36\% & 69.28\% & 51.80\%               \\ \hline
			$r_{V_0}$ &  $ \mathcal{G}^{V}\cup\mathcal{G}^{C} $  &
			52.59\% & 75.39\% & 78.82\% & 62.69\%
			              & 41.78\%     & 65.48\% & 68.80\% & 51.75\%                 \\ \hline
			$r_{V_1}$ & $\mathcal{G}^{V}\cup\mathcal{G}^{C}$   &
			51.73\% & 74.94\% & 78.22\% & 62.35\%
			               & 42.15\%     & 64.71\% & 68.19\% & 51.42\%
			                           \\ \hline
			$r_{V_2}$ & $\mathcal{G}^{V}\cup\mathcal{G}^{C}$   &
			52.05\% & 76.14\% & 79.33\% & 62.77\%
			              & 42.18\%     & 65.09\% & 68.74\% & 51.94\%                \\ \hline
			$r_{V_3}$ & $\mathcal{G}^{V}\cup\mathcal{G}^{C}$   &
			\textbf{53.11\%} & 76.31\% & 79.31\% & \underline{63.43\%}    & \underline{42.67\%}     & \underline{65.69\%} & \underline{69.29\%} & \underline{52.54\%}               \\ \hline
			$r_{V_4}$ & $\mathcal{G}^{V}\cup\mathcal{G}^{C}$   & \underline{53.09\%} & \textbf{76.75\% }& \underline{79.97\%} & \textbf{63.66}\%             & \textbf{42.91}\%     & \textbf{66.10}\% & \textbf{69.62}\% & \textbf{52.75}\%               \\ \hline
		\end{tabular}
		\caption{Performance of the mobility prediction based on the spatio-temporal mobility pattern relation ({\em STMPR}) with different auxiliary information, where bold denotes best (highest) results and underline denotes the second best results.}
	    \label{tab:result_aux}
	\end{table*}
	
\subsubsection{Impact of PoI category information} 
\begin{figure*}[t!]
\centering
\subfigure{\includegraphics[width=.40\textwidth]{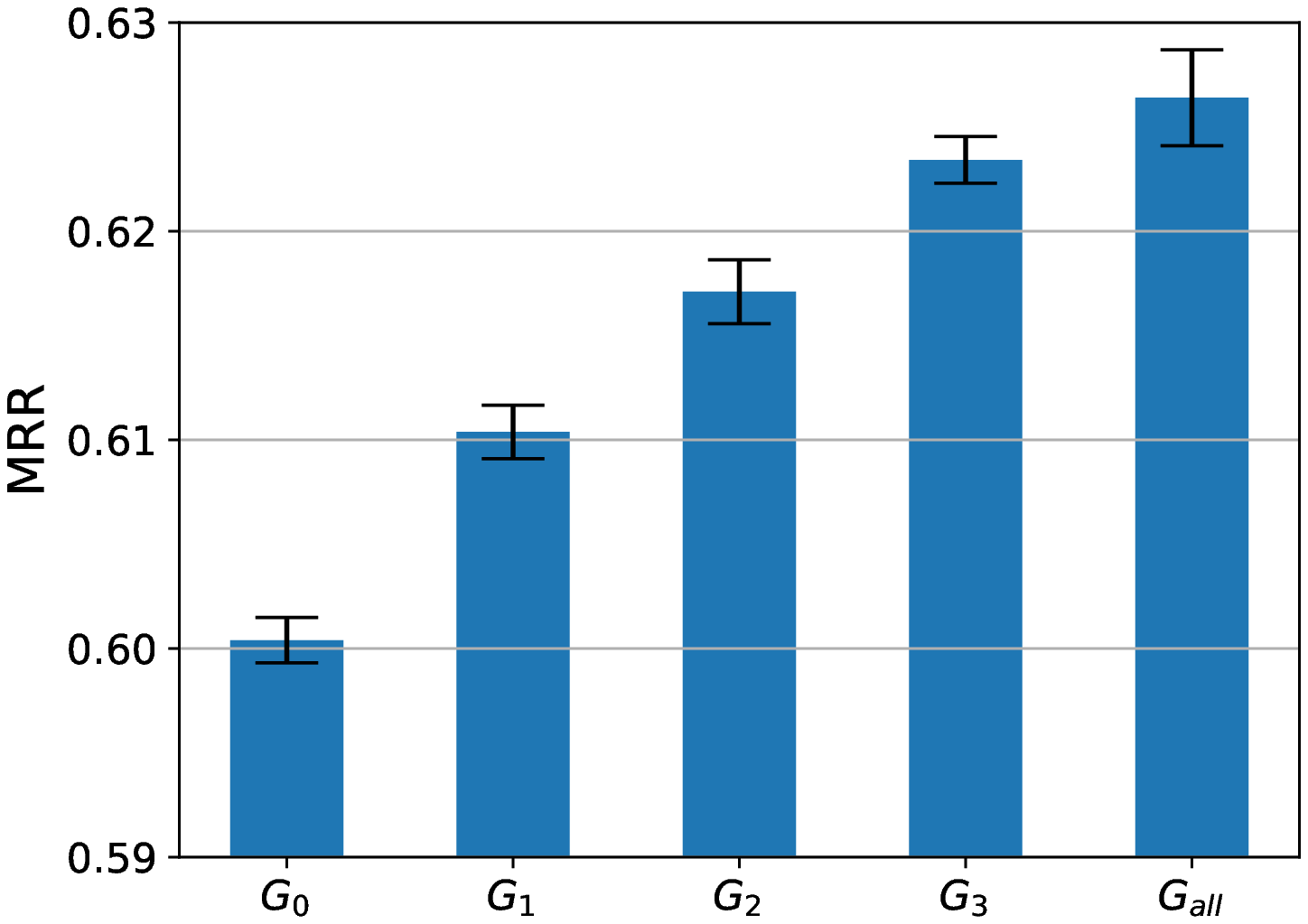}}
\subfigure{\includegraphics[width=.40\textwidth]{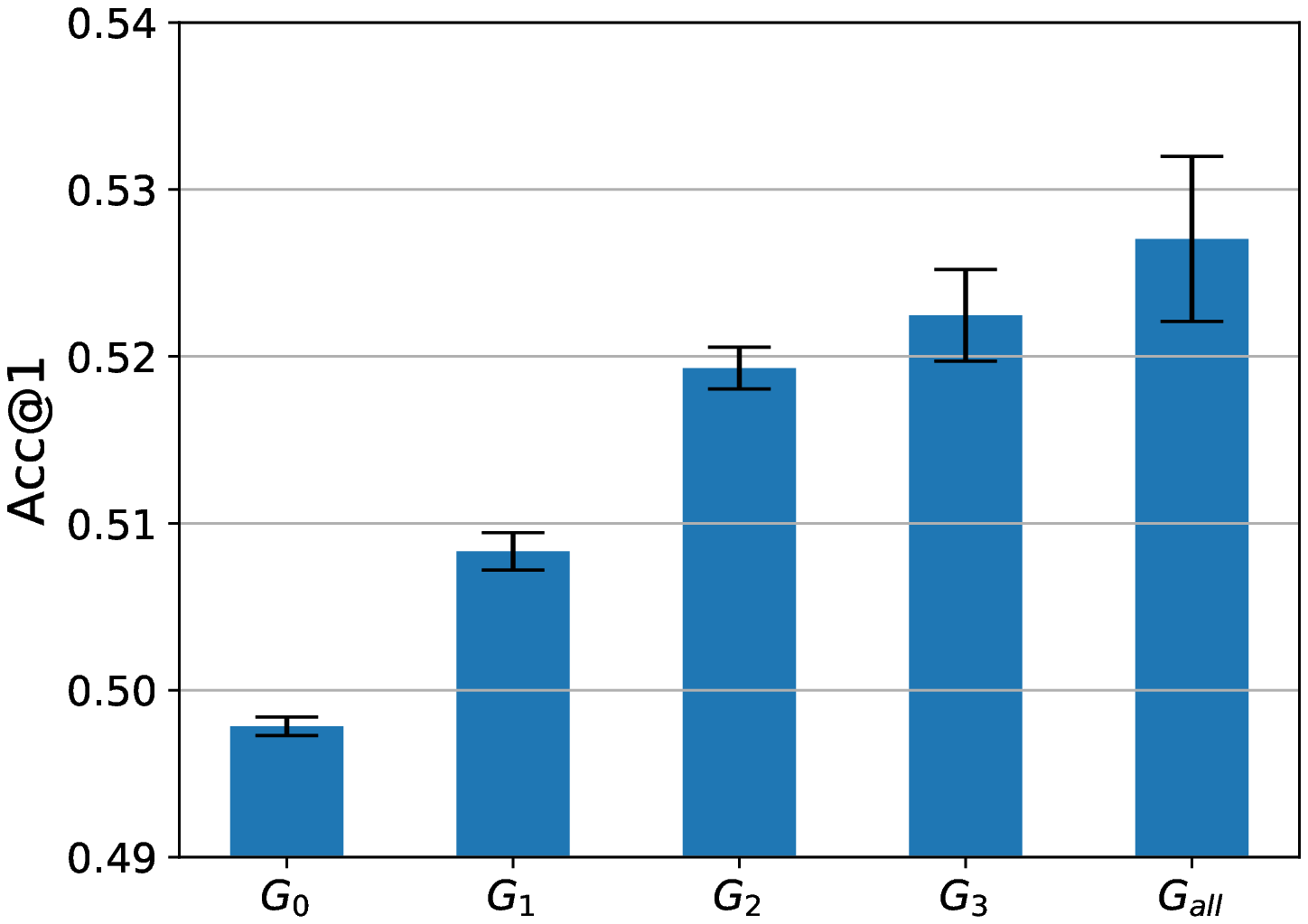}}
\caption{The performance of the models considering combining $\mathcal{G}^V$ and $\mathcal{G}^C$ on the Beijing dataset. Model $G_0$ means using $\mathcal{G}^V$ solely, Model $G_1$ means combining $\mathcal{G}^V$ with $\mathcal{G}^{C_1}$, Model $G_2$ means combining $\mathcal{G}^V$ with $\mathcal{G}^{C_2}$, Model $G_3$ means combining $\mathcal{G}^V$ with $\mathcal{G}^{C_3}$, Model $G_{all}$ means combining $\mathcal{G}^V$ with $\mathcal{G}^{C}$, i.e.,  $ \{\mathcal{G}^{C_1} \cup \mathcal{G}^{C_2} \cup \mathcal{G}^{C_3} \}$.}\label{fig:graph}
\end{figure*}

We then evaluate the performance of adding different PoI category information in our models.
Specifically, we show the performance of the models considering different combinations of $\mathcal{G}^V$ and $\mathcal{G}^C$ in Figure~\ref{fig:graph},
where the horizontal axis represents different models, while the vertical axis represents the values of MRR and Acc@$1$. $G_0$ $\sim$ $G_{all}$ represent our proposed methods with different combinations of $\mathcal{G}^V$ and $\mathcal{G}^C$. The left figure demonstrates that with the additional PoI category information, the performance of mobility prediction becomes better, i.e., the MRR is improved. Similarly, the results on the right figure also show the efficiency of adding the PoI category information to our models. Besides, combining different levels of category information achieves different performance gains, where leveraging coarse-level categories obtains more accurate results than utilizing fine-level categories. The main reason is that fine-level categories have more noise than other level categories. In general, the utilization of categories information indeed improves the whole performance of our proposed models.

\subsubsection{Impact of temporal dynamics}
\begin{figure*}[t]
\centering
\subfigure{\includegraphics[width=.75\textwidth]{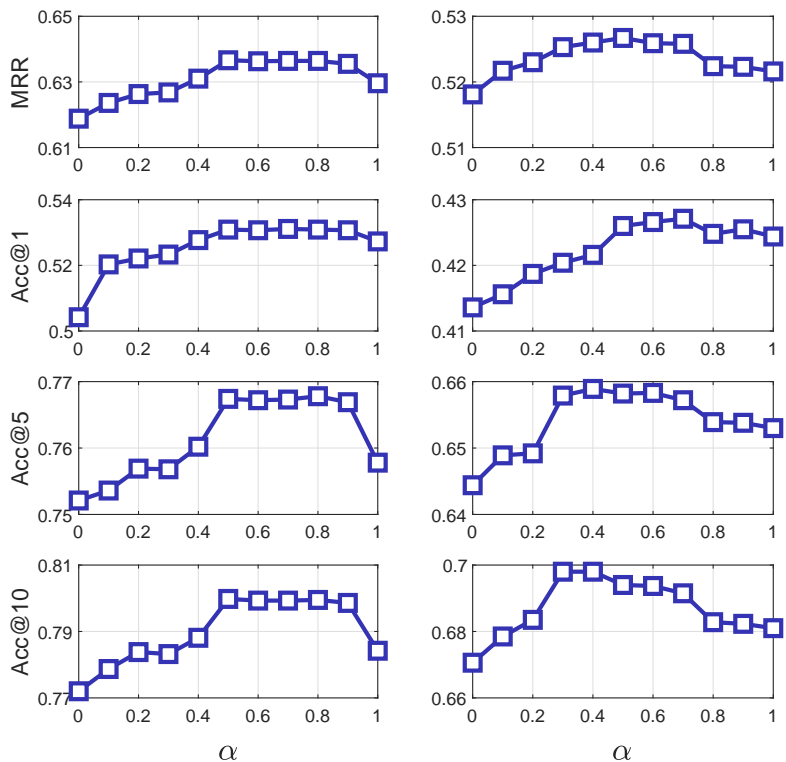}}
\caption{The performance on the Beijing dataset (left) and Shanghai dataset (right) as the function of the fraction of  $\alpha$. } \label{fig:alpha}
\end{figure*}

In order to interpret the influence of the temporal dynamic ratio $\alpha$, we show the mobility prediction performance of our proposed model in terms of different metrics as the functions of $\alpha$ in Figure~\ref{fig:alpha}.
The results on the left column are conducted on the Beijing dataset, and we can observe that when $\alpha$ is set to 0, the performance of our model becomes much worse. Besides, with the increase of $\alpha$, MRR and Acc@$K$ will reach the peak value first (about $\alpha=0.5$) and then begin to decrease, which is identical with our hypothesis and design.

\subsubsection{Impact of sampling rate $\&$ frequency of location access}
In this group of experiments, we discretize the timestamps of users’ trajectories into two-hour, one-hour, and half-hour time bins considering different sampling rates on the Beijing dataset to construct the {\em STKG}, respectively. Specifically, further distinguishing time bins belonging to working day and non-working day, we can obtain 24, 48, and 96 time bins, respectively. From Table~\ref{tab:sampling}, we can observe that our proposed method with half-hour time bins achieves the best performance, which suggests that the model with fine-grained trajectories performs better on mobility prediction.

\begin{figure*}[t!]
\centering
\subfigure{\includegraphics[width=.40\textwidth]{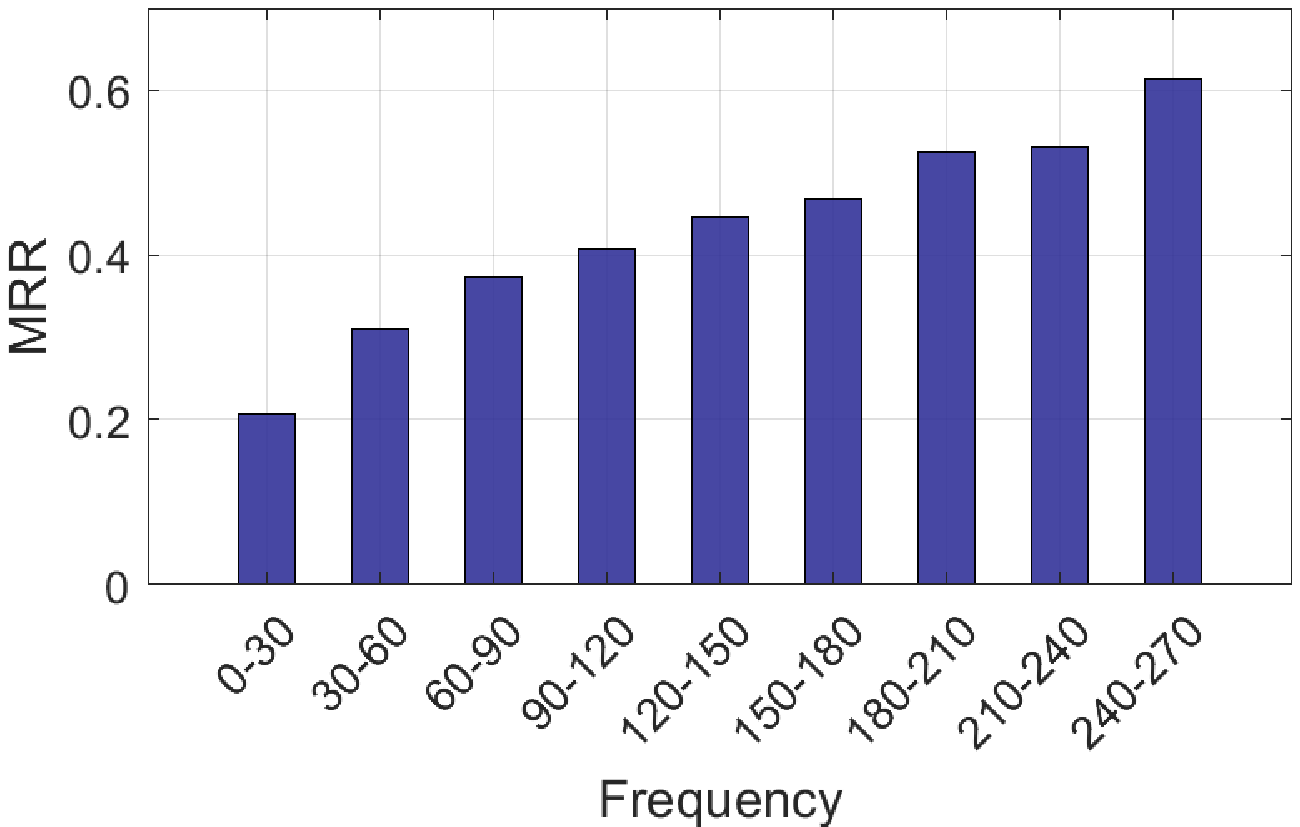}}
\subfigure{\includegraphics[width=.40\textwidth]{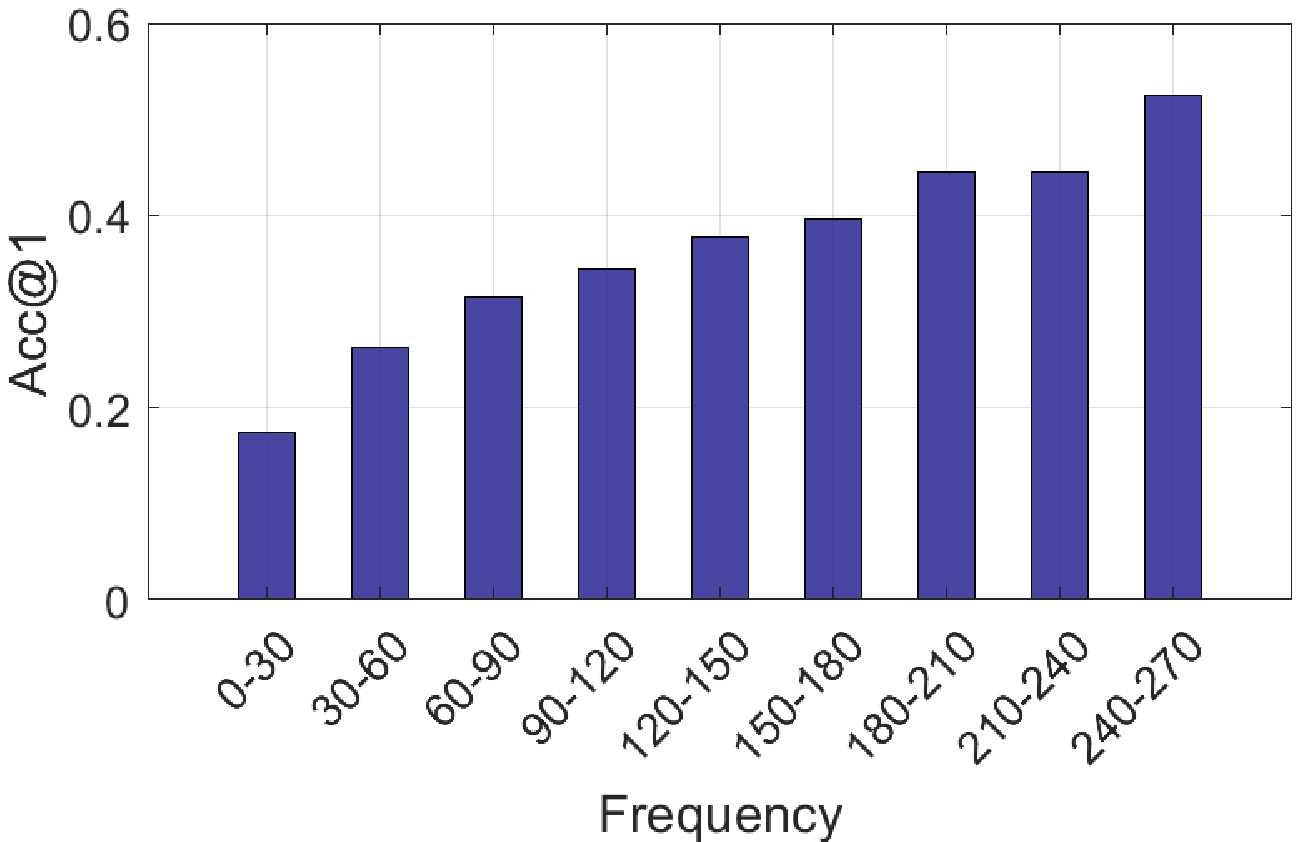}}
\subfigure{\includegraphics[width=.40\textwidth]{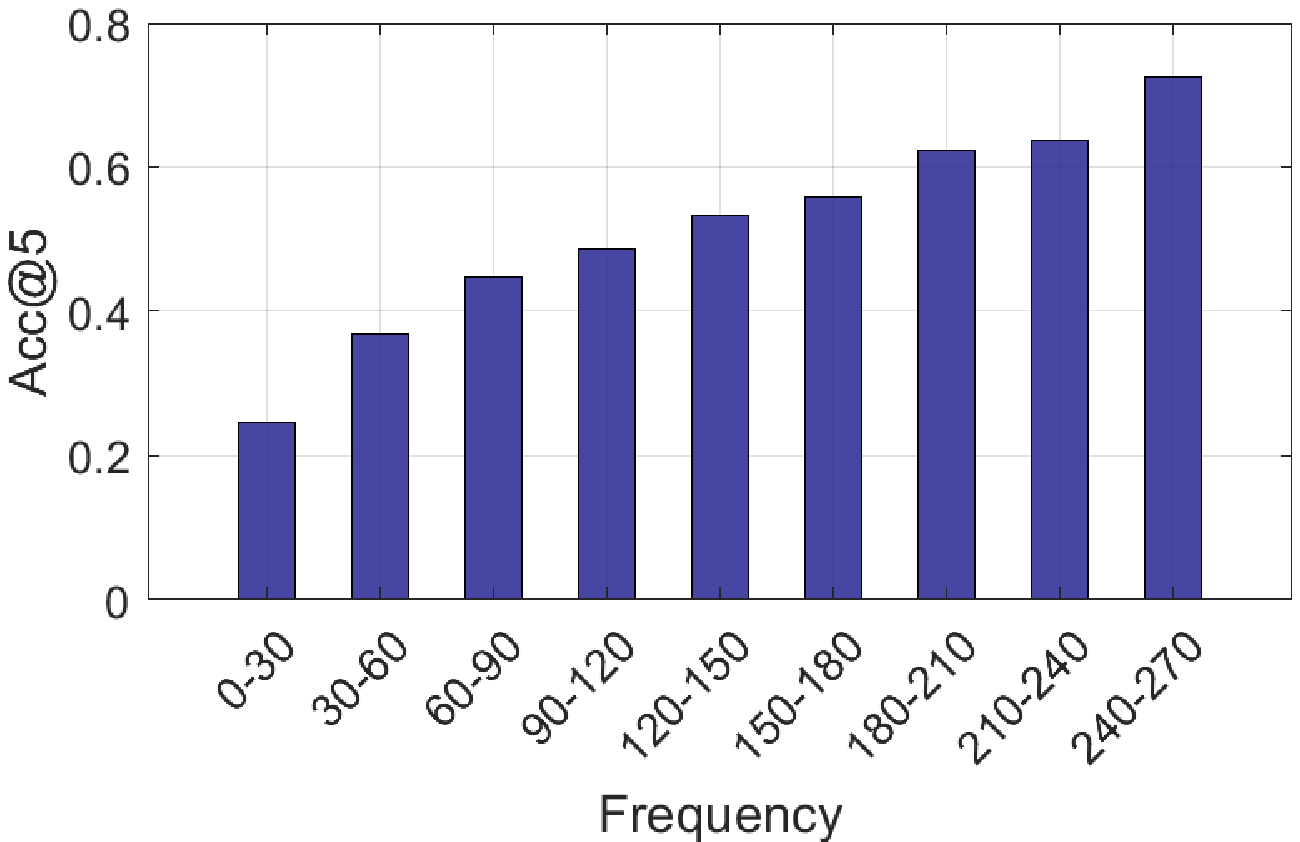}}
\subfigure{\includegraphics[width=.40\textwidth]{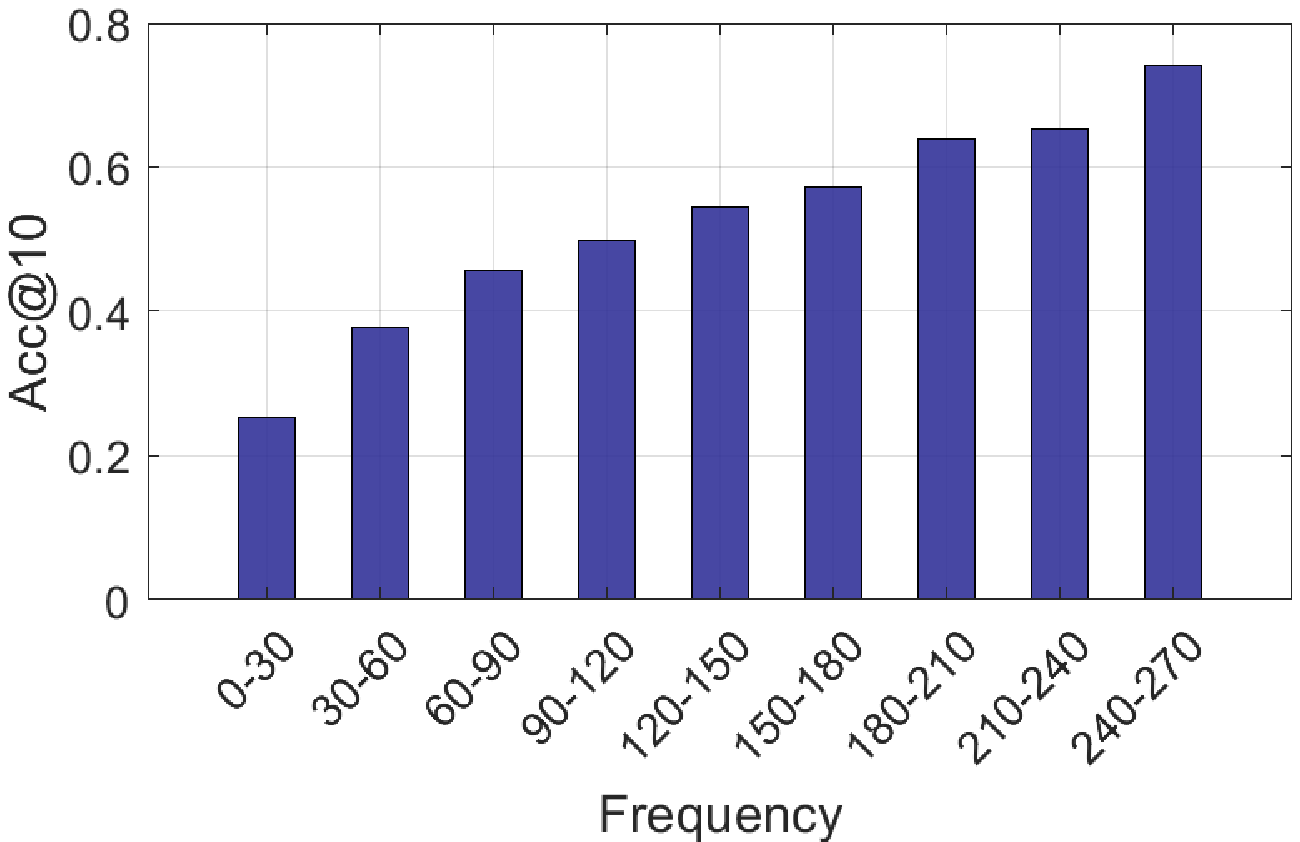}}
\caption{The mobility prediction performance of our proposed model as the functions of the frequency of user's location access on the Beijing dataset.}\label{fig:frequency}
\end{figure*}

We further divide PoIs into groups with different frequencies of user's location access, and show the average prediction performance corresponding to these PoI groups in Figure~\ref{fig:frequency}.
Specifically, the horizontal axis represents the visited frequency of PoIs. For example, the result corresponding to the frequency interval $30 \sim 60$ represents the average prediction accuracy of the locations which are visited by users for $30 \sim 60$ times in the historical mobility records. Generally speaking, the prediction is easier when users visit the same place more frequently, and we can conclude that with the increase of the frequency of user's location access, we can obtain a better prediction performance.

\subsubsection{Impact of other potential relations}
As we discussed before, the category affiliation relations are also possible to be considered in our constructed {\em STKG}.
Thus, in this group of experiments, we conduct comparative experiments on whether to add the category affiliation relation in our model, of which the results are shown in Table~\ref{tab:category_rel}.
As we can observe, considering the category affiliation relations indeed help improve the mobility prediction performance, demonstrating the effectiveness of our model in terms of capturing the rich semantic information from multiple types of relations. However, we can also observe that the performance gain is very limited. Thus, it is reasonable to mainly focus on the two key relationships.  Apart from the category affiliation relation, human mobility patterns may be affected by other relations, e.g., user feature, user social network, etc., which we leave for future work.

\begin{table*}
        \centering
		\begin{tabular}{|l||l|l|l|l|}
			\hline
			~         & \multicolumn{4}{c|}{Beijing} \\ \hline
			~         & \multicolumn{3}{c|}{Accuracy@k} & \multirow{2}{*}{MRR}    \\ \cline{1-4}
			~         & k=1        & k=5    & k=10    &                                    \\ \hline
			2 hours   & 52.22\% & 76.21\% & 79.42\% & 62.88\%     \\ \hline
		    1 hour   & 52.81\% & 76.62\% & 79.77\% & 63.38\%       \\ \hline			
			30 minutes   & 53.09 \% & 76.75\% & 79.97\% & 63.66\%      \\ \hline
		\end{tabular}
		\caption{The effects of sampling rate in terms of four metrics on the Beijing dataset.}
	    \label{tab:sampling}
	\end{table*}
	
	\begin{table*}
        \centering
		\begin{tabular}{|l|l||l|l|l|l||l|l|l|l|}
			\hline
			\multirow{3}{*}{\em STMPR}    & \multirow{3}{*}{\em STKG}     & \multicolumn{4}{c||}{Beijing} & \multicolumn{4}{c|}{Shanghai} \\ \cline{3-10}
			~    & ~     & \multicolumn{3}{c|}{Accuracy@k} & \multirow{2}{*}{MRR}  & \multicolumn{3}{c|}{Accuracy@k} & \multirow{2}{*}{MRR}  \\ \cline{3-5}\cline{7-9}
			~    & ~     & k=1        & k=5    & k=10    &                     & k=1        & k=5    & k=10    &                     \\ \hline
			$r_{V_4}$ & $\mathcal{G}^{V}\cup\mathcal{G}^{C}$   & 53.09\% & 76.75\% & 79.97\% & 63.66\%              & 	42.91\%     & 66.10\% & 69.62\% & 52.75\%           \\ \hline
			$r_{V_4}$ & $\mathcal{G}^{V}\cup\mathcal{G}^{C}\cup\mathcal{G}^{C'}$   & 53.19\% & 76.73\% & 80.01\% & 63.72\%               & 43.12\%     & 66.07\% & 69.61\% & 52.97\%               \\ \hline
		\end{tabular}
		\caption{Performance of the mobility prediction based on the spatio-temporal mobility pattern relation ({\em STMPR}) with two types of affiliation relation.}
	    \label{tab:category_rel}
	\end{table*} 
\section{Related Work and Discussion}\label{sec:relatedwork}

\subsection{Mobility Prediction}

A number of works focus on predicting users' future mobility based on historical user trajectories directly.
One widely used method is the Markov models~\cite{lu2013approaching, Song2006Evaluating}.
In the Markov model, users' mobility is modeled by a series of transitions between the predefined states, which represent different locations.
In addition, users' future movements only depend on their current states.
Wang et al.~\cite{zeng2017predictability} utilize multiple interconnected Markov chains to model users' different mobility behavior during different periods of time.
Jeong et al.~\cite{jeong2016cluster} cluster users based on their Markov transition kernels.
Then, the obtained cluster structure is further used to aid mobility prediction and overcome the sparse issue.
An extension of the Markov model is the high-order Markov model, where users' future states depend on the past several states.
Lu et al.~\cite{lu2013approaching} combine multiple Markov models with different orders to predict users' mobility, where a ``fallback mechanism'' is adopted to dynamically switch between Markov models with different orders \cite{Song2006Evaluating}.
Similar with~\cite{lu2013approaching}, a variable-order Markov model with the dynamic order adaptively determined by users' personal frequent patterns is also utilized in~\cite{qiao2018hybrid} to predict users' mobility.
Another extension of the Markov model is the hidden Markov model (HMM), where each state represents a probabilistic distribution over locations rather than a single location.
Mathew et al.~\cite{Mathew2012Predicting} utilize HMM to predict users' mobility.
Zhang et al.~\cite{zhang2016gmove} further combine HMM and clustering techniques~\cite{jeong2016cluster} to improve the mobility prediction performance.
Input-Output Hidden Markov Model (IO-HMM), which is another variation of the Markov model and can incorporate context information of the sequence to be predicted, is utilized in \cite{hu2015iohmm} to model the temporal information.
Different from the Markov model, a number of other approaches model users' mobility based on Bayesian probabilistic models.
Wang et al.~\cite{wang2020human} consider the continuity of human mobility in terms of the temporal dimension by utilizing von Mises distribution in the Bayesian probabilistic model.
Huai et al.~\cite{huai2014toward} extend the Bayesian hidden Markov model to consider contextual information.
Jie et al.~\cite{Jie2018DeepMove} utilize the recurrent neural network combined with the attention mechanism to model user mobility.
McInerney et al. \cite{mcinerney2013modelling} propose a Bayesian-based algorithm that shares temporal patterns of mobility among different users.
Compared with them, our work is the first to convert the mobility prediction problem to the fact completion problem in knowledge graphs. 
Compared with the existing method~\cite{GuoSunZhangTheng2020} which uses the embeddings extracted from the knowledge graph as the initial features, the technique of knowledge graphs is a more powerful tool and can incorporate background knowledge from multiple sources and multiple types of relations in a cohesive manner, which have been demonstrated by our experimental results.
What's more, the ``knowledge'' involved in the massive spatio-temporal trajectories of users can be extracted in the form of structured tuples, which help to better understand the user mobility patterns and better characterize the features of users, spatial venues, and time correlated with the mobility patterns of users.

\subsection{Knowledge Graph}
Traditionally, knowledge graph (KG) represents real-world facts with relations and corresponding involved entities~\cite{wang2017knowledge, ji2021survey}. 
Especially, studies close to our work are spatial KGs \cite{sun2019demonstrating,yan2019spatially,Wang2020Incremental} and temporal KGs \cite{trivedi2017know,lacroix2020tensor,goel2020diachronic,dasgupta2018hyte}. 
As for spatial KGs, most works investigate the construction with explicit geographic entities. 
Sun et al. \cite{sun2019demonstrating} utilize the taxonomy tree structure to describe the spatial relationship between geographic entities, e.g., China isPartOf Asia. 
Yan et al. \cite{yan2019spatially} propose a reinforcement learning framework to explicitly summarize geographic components such as places and regions. 
Recently, Wang et al. \cite{Wang2020Incremental} construct a spatial KG for mobile user profiling, in which regions, PoIs, and categories are defined as entities, while isPartOf and isCategoryOf are defined as relations. 
However, such KGs above are static without time domain considered, which usually fails in spatio-temporal modeling. 
On the other hand, temporal KGs incorporate time information into traditional KGs with quadruplets, i.e., the relationship between head and tail entities as well as time points. 
Several works focus on temporal KG completion. 
For example, Trivedi et al. \cite{trivedi2017know} firstly integrate temporal point process with recurrent neural network for temporal KG completion. 
Goel et al. \cite{goel2020diachronic} consider both static and dynamic features of entities, and further propose diachronic entity embedding for temporal KG completion. 
In addition, Dasgupta et al. \cite{dasgupta2018hyte} follow translation distance approaches in static KGs, leveraging time-related hyperplane for semantic modeling. Lacroix et al. \cite{lacroix2020tensor} extend tensor decomposition approach ComplEx \cite{Trouillon2016Complex} from static KGs to temporal KGs with complex-valued time embedding proposed. 
It should be noted that existing temporal KG completion approaches only investigate event based KGs, especially political events with the coarse temporal granularity of 1 day~\cite{leetaru2013gdelt,ward2013comparing}. 
In comparison, our work focus on the fine-grained temporal granularity of half an hour and much more complex human mobility problems. 
Additionally, Neumaier et al. \cite{NEUMAIER201921} integrate the relevant tables through leveraging spatio-temporal annotations and construct a hierarchical spatio-temporal knowledge graph to capture the semantic information between spatial entities, e.g., the geo-entities federal state `Bavaria' locates in the country `Germany'. Different from them, our work directly defines the affiliation relations characterizing PoI category information to assist in improving the performance of the mobility prediction.
Overall, existing works in KGs consider either spatial domain or time domain with quite coarse spatial granularity (city, country, etc.) or temporal granularity (day, year, etc.), which are not applicable for fine-grained human mobility prediction (usually PoI and hour for spatial and temporal granularity). In comparison, our work considers both spatial and temporal domains for such problems, which helps us to model the mobility patterns of users in terms of both temporal dimension and spatial dimension and improve the mobility prediction performance.
Besides, several recent works observe that real-world knowledge usually involves more than two entities simultaneously (like our researched problem), and generalize traditional binary KGs to $n$-ary KGs with $n$-ary relations~\cite{liu2020generalizing,guan2019link}, which can be good support for our research. Finally, by using the spatio-temporal context information as the auxiliary information, the mobility prediction performance is significantly improved.


\subsection{Implications and Limitations}

Our study has shown that our defined {\em STKG} and the involved spatio-temporal mobility pattern relation are an effective carrier of ``knowledge'' involved in the massive mobility trajectory data. Specifically, based on them, the performance of mobility prediction has beat the most representative mobility state-of-the-art algorithms. 
What's more, our study has shown that more facts as background knowledge help to better model the facts of the spatio-temporal mobility pattern relations, indicating that we can characterize the world better by introducing more knowledge into the knowledge graph. A more important implication for us is that the knowledge graph technique can be utilized to solve many ubiquitous computing tasks, where our work has made a good example. What's more, based on the strong ability of the knowledge graph in terms of incorporating ``knowledge” from multiple sources and multiple types of relations, a universal knowledge graph, which incorporates all the available data sources or relation types, is possible to be built to solve massive ubiquitous computing tasks simultaneously, and even achieve better performance by utilizing the correlation between different tasks. For the community of the computer human interaction, the knowledge graph can also help to express the ``knowledge'' about the interaction between the computer and human better, and provide a new promising solution to solve many practical tasks. For example, some research variables and their relationships in both ubiquitous computing and computer human interaction can be seen as the entities and relations in the knowledge graph, and widely studied knowledge graph embedding approaches \cite{wang2017knowledge} can provide good representations for such variables in downstream tasks.
In addition, the extracted ``knowledge'' in the form of structured tuples is also useful for a number of applications including user profiling~\cite{xu2018detecting}, spatio-temporal aware recommendation~\cite{yuan2017pred}, etc.

One of the limitations of our paper is that we only consider the background knowledge of the PoI categories, which is one of the most readily available background knowledge and can be obtained from online map services. 
Future work will look into the impact of background knowledge from different sources, e.g., fine-grained attribute information of spatial venues and users, spatio-temporal event information in the urban environment, etc.
Another limitation is that the temporal information is only utilized in the form of discrete time bins in our paper. There exist other techniques of knowledge graph embedding that utilize the continuous temporal information, e.g.,  diachronic embedding~\cite{goel2020diachronic}. Future work will look into the impact of different forms of temporal information on the mobility prediction task.

\section{Conclusions and Future work}\label{sec:conclusion}
In this paper, we propose a new type of knowledge graph, i.e., spatio-temporal urban knowledge graph ({\em STKG}), where mobility trajectories are modeled as the facts of spatio-temporal mobility pattern relations, and we convert the mobility prediction problem to the knowledge graph completion problem in the constructed {\em STKG}. Further, combined with the background knowledge from other sources, e.g., Point of Interests (PoI) category information, a complex embedding model is trained to learn the features of entities and relations and measure the plausibility of facts in {\em STKG} to solve the knowledge graph completion problem. Extensive evaluations confirm the high accuracy of our model in predicting users' mobility and demonstrate the effectiveness of utilizing PoI categories as the background knowledge in {\em STKG}. As for future work, we will consider more background knowledge from different sources. In addition, we will consider multiple variants of the spatio-temporal mobility pattern in one {\em STKG} simultaneously, and utilize the structure of relation paths in {\em STKG} to improve the prediction performance.

\begin{acks}
This work was supported in part by
the National Natural Science Foundation of China under U1936217,  61971267, 61972223, 61941117, 61861136003, and 62171260,
the National Key Research and Development Program of China under grant 2018YFB1800804,
the Beijing Natural Science Foundation under L182038,
the Beijing National Research Center for Information Science and Technology under 20031887521,
the China Postdoctoral Science Foundation under 2021M691830,
and research fund of Tsinghua University - Tencent Joint Laboratory for Internet Innovation Technology.
\end{acks}

\bibliographystyle{ACM-Reference-Format}
\bibliography{Reference,RefShort}

\end{document}